\documentclass[letterpaper]{emulateapj}

\usepackage{epsfig}
\usepackage{amsmath}
\usepackage{epsfig,rotating}
\usepackage{natbib}
\usepackage{lscape}
\usepackage{enumerate}
\usepackage{multirow}
\usepackage{array}
\def\sles{\lower2pt\hbox{$\buildrel {\scriptstyle <}
   \over {\scriptstyle\sim}$}}

\def\sgreat{\lower2pt\hbox{$\buildrel {\scriptstyle >}
   \over {\scriptstyle\sim}$}}
\newcommand{\cN}[1]{\mathcal{N}}

\def\gsim{\;\rlap{\lower 2.5pt
 \hbox{$\sim$}}\raise 1.5pt\hbox{$>$}\;}
\def\lsim{\;\rlap{\lower 2.5pt
   \hbox{$\sim$}}\raise 1.5pt\hbox{$<$}\;}

\slugcomment{Accepted for publication in ApJ, March 8, 2010}

\begin{document}


\title{Tidal Heating Models for the Radii of the Inflated Transiting
       Giant Planets WASP-4\lowercase{b}, WASP-6\lowercase{b}, WASP-12\lowercase{b}, WASP-15\lowercase{b}, and TrES-4}

\author{Laurent Ibgui\altaffilmark{1}, Adam Burrows\altaffilmark{1}, and David S. Spiegel\altaffilmark{1} }

\affil{$^1$Department of Astrophysical Sciences, Peyton Hall, Princeton University, Princeton, NJ 08544}

\vspace{0.5\baselineskip}

\email{ibgui@astro.princeton.edu, burrows@astro.princeton.edu, dsp@astro.princeton.edu}

\begin{abstract}
In order to explain the inflated radii of some transiting extrasolar
giant planets, we investigate a tidal heating scenario for the
inflated planets WASP-4b, WASP-6b, WASP-12b, WASP-15b, and TrES-4.  To
do so, we assume that they retain a nonzero eccentricity, possibly by
dint of continuing interaction with a third body. We calculate the
amount of extra heating in the envelope that is then required to fit
the radius of each planet, and we explore how this additional power
depends on the planetary atmospheric opacity and on the mass of a
heavy-element central core. There is a degeneracy between the core
mass $M_{\rm core}$ and the heating $\dot{E}_{\rm
  heating}$. Therefore, in the case of tidal heating, there is for
each planet a range of the couple $\{M_{\rm core},e^2/Q'_p\}$ that can
lead to the same radius, where $Q'_{p}$ is the tidal dissipation
factor and $e$ is the eccentricity. With this in mind, we also
investigate the case of the non-inflated planet HAT-P-12b, which can
admit solutions combining a heavy-element core and tidal heating.  A
substantial improvement of the measured eccentricities of such
planetary systems could simplify this degeneracy by linking the two
unknown parameters $\{M_{\rm core},Q'_{p}\}$. Further independent
constraints on either of these parameters would, through our
calculations, constrain the other.
\end{abstract}


\section{Introduction}
\label{sec:intro}
The transiting extrasolar giant planets (EGPs), around 60 discovered
to date\footnote{See J. Schneider's Extrasolar Planet Encyclopaedia at
  http://exoplanet.eu, the Geneva Search Programme at
  http://exoplanets.eu, and the Carnegie/California compilation at
  http://exoplanets.org.}, offer the best testbed for theoretical
models of the evolution of planetary radii.  In the last 14 years,
many such theoretical models and tests have been proposed and
investigated \citep{guillot_et_al1996, burrows_et_al2000,
  Bodenheimer_et_al_2001, burrows_et_al2003, Baraffe_et_al_2003,
  Bodenheimer_et_al_2003, Gu_et_al_2003,burrows_et_al2004,
  Fortney_and_Hubbard_2004, Baraffe_et_al_2004, Chabrier_et_al_2004,
  laughlin_et_al_2005_1, Baraffe_et_al_2005, Baraffe_et_al_2006,
  burrows_et_al2007, Fortney_et_al_2007, Marley_et_al_2007,
  chabrier+baraffe2007, Liu_et_al_2008, Baraffe_et_al_2008,
  Ibgui_and_Burrows_2009, Miller_et_al_2009, Leconte_et_al_2009}.

A gas giant planet's radius is a function of many variables, including
the planet's mass and age; the stellar irradiation flux and spectrum;
the composition -- in particular, the heavy-element content -- of the
atmosphere, the envelope, and the core; the atmospheric circulation
that couples the day and the night sides; and any processes that could
generate an extra power source in the interior of the planet, such as
tidal heating.  Furthermore, the connection between a planet's
physical radius and its transit radius is complicated by the transit
radius effect \citep{burrows_et_al2003, Baraffe_et_al_2003}.  Since
each of these variables can change from one planet-star system to
another, only a custom calculation can determine if the measured
transit radius of a giant planet matches theoretical predictions.

The first observations of a transiting planet (HD209458b), which found
that its diameter is more than 30\% greater than Jupiter's
\citep{henry_et_al2000, charbonneau_et_al2000}, revealed a gap in our
understanding of radius evolution. HD209458b's radius is significantly
larger than the standard evolutionary theory would predict for an
object of its age \citep{knutson_et_al2007a, burrows_et_al2007}.  For
this reason it is said to be inflated, or bloated.  In the last
decade, a number of other inflated planets have been discovered, whose
radii are similarly difficult to explain in terms of simple
evolutionary theory (CoRoT-1b: \citealt{Barge_et_al_2008,
  Gillon_et_al_2009_3}, CoRoT-2b: \citealt{Alonso_et_al_2008},
HAT-P-1b: \citealt{bakos_et_al2007a, Winn_et_al_2007,
  Johnson_et_al_2008}, TrES-2: \citealt{odonovan_et_al2006b}, TrES-4:
\citealt{Mandushev_et_al_2007, Sozzetti_et_al_2008}, WASP-4b:
\citealt{Wilson_et_al_2008, Southworth_et_al_2009_2,
  Gillon_et_al_2009_1, Winn_et_al_2009_1}, WASP-6b:
\citealt{Gillon_et_al_2009_2}, WASP-12b: \citealt{Hebb_et_al_2009},
WASP-15b: \citealt{West_et_al_2009}, XO-3b:
\citealt{Johns-Krull_et_al_2008, Winn_et_al_2008}).

The present paper investigates a stationary heating scenario as a
possible explanation of the inflated transiting giant planets recently
discovered: WASP-4b, WASP-6b, WASP-12b, and WASP-15b.  In addition, we
revise the case of TrES-4, already discussed by
\citet{Liu_et_al_2008}, and consider HAT-P-12b
\citep{Hartman_et_al_2009}, a planet that is not inflated, but that
appears to require a massive, dense core to explain its small transit
radius. At the time of the writing of this paper, three other inflated
transiting EGPs have been discovered: CoRoT-5b
\citep{Rauer_et_al_2009}, HAT-P-13b \citep{Bakos_et_al_2009_2}, and
WASP-17b \citep{Anderson_et_al_2009}.  Although we might consider them
in detail in a future study, we find it reassuring that preliminary
investigations of these objects indicate that the general philosophy
of the present paper can be applied to them as well.

Tidal heating as an explanation for inflated close-in EGPs was
proposed by \citet{Bodenheimer_et_al_2001, Bodenheimer_et_al_2003}.
This heating being due to a nonzero eccentricity, they suggested the
necessity of an excitation mechanism, as for example gravitational
interaction with another planetary companion.  This possibility was
examined by \citet{Mardling_2007}.  \citet{Liu_et_al_2008}
investigated the cases of the inflated planets TrES-4, XO-3b, and
HAT-P-1b, while assuming a constant rate of interior heating.
\citet{Ibgui_and_Burrows_2009}, \citet{Ibgui_et_al_2009_2}, and
\citet{Miller_et_al_2009} coupled the evolution of the planetary
radius with that of the orbit, under the influence of tidal
interactions.  The interesting feature revealed by this scenario is a
possible transient inflation of the planetary radius that temporarily
interrupts its monotonic standard shrinking.

Here, we assume a constant heating rate. Such an assumption would be
valid, in the context of tidal heating, if there is a stationary
scenario such that the orbital eccentricity and semimajor axis of the
transiting EGP are quasi-constant and the radius has reached a
``quasi-equilibrium'' value.  This paper is organized as follows. In
Section~\ref{sec:properties}, we present the relevant properties of
the planet-star systems in our study.  In Section~\ref{sec:model}, we
briefly summarize the model assumptions adopted for our planetary
radius evolutionary calculations.  In Section~\ref{sec:theo_Req}, we
present both the details of our extra-heating model and its
results. We derive the necessary heating rates $\dot{E}_{\rm heating}$
for the equilibrium radii $R_{\rm eq}$ to match the measured values.
In Section \ref{subsec:no_central_core}, we assume that the planets
have no heavy-element central core and show that an enhanced planet
atmospheric opacity, since this results in a larger radius, requires
less internal heating.  In Section \ref{subsec:central_core_solar}, we
freeze the planetary atmospheric opacity, and show that the presence
of a heavy-element central core, whose effect is to shrink the
planetary radius, requires more internal heating.  In Sections
\ref{subsec:no_central_core} and \ref{subsec:central_core_solar}, we
relate $\dot{E}_{\rm heating}$ to the ratio $e^2/Q'_p$, where $Q'_p$
is the tidal dissipation factor \citep{Goldreich_1963} and $e$ the
orbital eccentricity.  In Section \ref{subsec:compatible_models} we
further show the following degeneracy: a given radius can be matched
by a range of values of the couples $\{M_{\rm core},\dot{E}_{\rm
  heating}\}$ or $\{M_{\rm core},e^2/Q'_p\}$. When the measured
eccentricities become sufficiently reliable, we could directly relate
$M_{\rm core}$ and $Q'_p$. We make a first attempt at such a relation
in this paper, while we remain fully aware of the poor constraints on
$e$.  We conclude and provide a general discussion in Section
\ref{sec:conclusion}.

\section{Observed Properties of the Systems under Consideration}
\label{sec:properties}
We consider the inflated planets WASP-4b, WASP-6b, WASP-12b, WASP-15b,
and TrES-4, and the non-inflated planet HAT-P-12b.  The relevant
parameters for our study are listed in
Tables~\ref{tab:transit_planets_data} (planets' properties) and
\ref{tab:host_stars_data} (host stars' properties).

For each observational study, the status of these planets, inflated or
non-inflated, was determined by comparing the observed radius to
tabulated generic radii \citep{Bodenheimer_et_al_2003,
  Fortney_et_al_2007, Baraffe_et_al_2008}. This procedure may be too
restrictive, given the multiple variables on which the radius depends
-- the planet's mass and age, the stellar irradiation flux and
spectrum, the atmospheric composition, the presence of heavy elements
in the envelope or in a central core, the atmospheric circulation that
couples the day and the night sides, and any effects that could
generate an extra power source in the interior of the planet, such as
tidal heating.  Only a calculation that is customized to the actual
planet-star system can provide the most reliable prediction of the
transit radius of the planet.

WASP-4b was discovered by \citet{Wilson_et_al_2008} and its parameters
were further refined by \citep{Southworth_et_al_2009_2,
  Gillon_et_al_2009_1, Winn_et_al_2009_1}.  It has an observed radius
of $1.365^{+0.021}_{-0.021}~R_{J}$ with a mass of
$1.237^{+0.064}_{-0.064}~M_{J}$ for an age of
$6.5^{+2.3}_{-2.3}$~Gyr. Its eccentricity is not well constrained, but
\citet{Madhusudhan_and_Winn_2009} derived an upper limit of 0.096.

WASP-6b was discovered by \citet{Gillon_et_al_2009_2}, who observe a
radius of $1.224^{+0.051}_{-0.052}~R_{J}$, and a small mass of
$0.503^{+0.019}_{-0.038}~M_{J}$.  They derive a nonzero eccentricity
of $0.054^{+0.018}_{-0.015}$. \citet{Gillon_et_al_2009_2} categorize
the planet as clearly inflated. However, as we show later, our
calculation indicates that the theoretical radius fits the measurement
within the 1$\sigma$ limit, without needing to invoke extra
heating. It might be a slightly inflated planet; more precise
measurements in the future might resolve whether extra heating needs
to be invoked to explain its size.

WASP-12b was discovered by \citet{Hebb_et_al_2009}. It has several
singular properties.  It is the second largest transiting planet
discovered to-date with a radius of $1.79 ^{+0.09} _{-0.09}~R_{J}$
(WASP-17b, discovered by \citealt{Anderson_et_al_2009}, is the largest
one with a radius of $1.994 ^{+0.096} _{-0.099}~R_{J}$). Its mass is
$1.41^{+0.10 }_{-0.10 }~M_{J}$. Its estimated eccentricity is
$e=0.049^{+0.015}_{-0.015}$.  It is also the most heavily irradiated
transiting EGP, with a flux at the substellar point of 9.098$\rm
\times 10^{9}~ergs~cm^{-2}~s^{-1}$. This planet has one of the
shortest orbital periods, $P=1.09142 ~\rm days$. Finally, and perhaps
most interesting, its orbital separation is close to its estimated
Roche limit, 0.0221~AU, while its periastron $p=a(1-e)$ might even be
smaller than this limit, depending upon the true values of its
semimajor axis and eccentricity. The planet is, therefore, perhaps on
the verge of being tidally disrupted. It may also be losing mass by
exceeding its Roche lobe \citep{Gu_et_al_2003, Li_et_al_2009}.  We do
not take this process into account in our model.
\citet{Miller_et_al_2009} suggest a floor on eccentricity in order to
fit the radius.

WASP-15b was discovered by \citet{West_et_al_2009}. Its observed
radius is $1.428^{+0.077}_{-0.077}~R_{J}$; its mass is
$0.542^{+0.050}_{-0.050}~M_{J}$, and its eccentricity is
$0.052^{+0.029}_{-0.040}$.

TrES-4 was discovered by \citet{Mandushev_et_al_2007}.  Its latest
updated parameters are from \citep{Sozzetti_et_al_2008}.  Its observed
radius is $1.783^{+0.093}_{-0.086}~R_{J}$, and its mass is
$0.925^{+0.081}_{-0.082}~M_{J}$.  As for the eccentricity,
\citet{Knutson_et_al_2009_1} report a $3 \sigma$ upper limit for
$|e\cos(\omega)|$ of 0.0058, where $\omega$ is the argument of
periastron.  We adopt a maximum value of 0.01.

HAT-P-12b, discovered by \citet{Hartman_et_al_2009} is not an inflated
planet. It has a small mass $0.211^{+0.012}_{-0.012}~M_{J}$ for a
radius of $0.959^{+0.029}_{-0.021}~R_{J}$. It has an upper limit on
eccentricity of $0.065$. As we will see later, our models require a
heavy-element core to fit its radius at its estimated age.

\section{Evolutionary Model Assumptions}
\label{sec:model}
The giant planet radius evolutionary model has been discussed in
detail in \citet{burrows_et_al2003}, \citet{burrows_et_al2007}, and
\citet{Ibgui_and_Burrows_2009}.  We briefly summarize its principal
features.

We model the evolution, with age, of the radii of giant planets with
the Henyey code of \citet{Burrows_et_al_1993, Burrows_et_al_1997}.
The boundary conditions \citep{burrows_et_al2003} incorporate
realistic irradiated planetary atmospheres calculated by CoolTLUSTY, a
variant of TLUSTY \citep{Hubeny_1988, Hubeny_and_Lanz_1995}.  In some
circumstances, the presence of an extra upper-atmosphere absorber can
lead to radiative equilibrium solutions that have thermal inversions
in which, above a relative minimum, atmospheric temperatures increase
with height \citep{hubeny_et_al2003, Burrows_et_al_2008_1,
  Burrows_et_al_2008_2, Knutson_et_al_2008, Fortney_et_al_2008_1,
  Spiegel_et_al_2009_1}.  However, in this study, we do not consider
such thermal inversion cases. The host star spectrum is calculated by
interpolation at its actual effective temperature and gravity, from
the \citet{Kurucz_1994} models. Such a customized approach should in
principle enhance the reliability of the model, in comparison with
generic and pre-calculated planetary atmosphere tables.  The planet
structure consists of an isentropic (convective) gaseous envelope and
a possible inner heavy-element core. The envelope consists of a
mixture of H$_{2}$ and He (helium mass fraction $Y=0.25$), and is
described by the equation of state of \citet{Saumon_et_al_1995}.  The
possible core (evidence for which is described in \citealt{guillot06,
  burrows_et_al2007, Guillot_2008}) is described by the ANEOS equation
of state of \citet{Thompson_and_Lauson_1972}.

The extra power source is assumed to be entirely in the interior of
the planet, i.e. in its convective zone.  We do not include any
heating in the radiative region.  Moreover, the heating rate is
assumed to be constant, as a proxy for the more complicated behavior
that actually might actually result if a companion excites the
eccentricity of the transiting planet.  This assumption was made by
\citet{burrows_et_al2007} and \citet{Liu_et_al_2008}, who calculated
the rate of energy dissipation in several EGPs (HD209458b, TrES-4,
XO-3b, and HAT-P-1b) needed to maintain their radii at their observed
values.  With the hypothesis that this source comes from tidal
interactions, this configuration may result from the gravitational
influence of a second (or companion) planet that would pump up the
eccentricity of the transiting planet \citep{Bodenheimer_et_al_2001,
  Mardling_2007}, preventing it from circularizing.

Specifically, \citet{Mardling_2007} has demonstrated that, depending
on a system's architecture, an external pertuber could maintain the
eccentricity of a given transiting EGP at quasi-constant values over
gigayear timescales, or could excite the transiting planet's
eccentricity so that it oscillates around a mean value on a timescale
that is fast compared with the planet's Kelvin-Helmholtz time, so from
a tidal heating perspective the planet responds as though the heating
is essentially constant in time.  Reproducing Mardling's calculations
in the cases of the planets examined in the current paper is not our
focus here.  Only such calculations, performed on a case-by-case
basis, can provide constraints on the physical and orbital properties
of a possible companion capable of sustaining the orbital eccentricity
(and the semimajor axis) of a given transiting planet, and can
determine the contribution of this companion to the energy budget of
the system.  However, we note that it is, at the present, plausible
that such external perturbers might exist in the systems under
consideration herein, for two reasons:
\begin{itemize}
\item First, to date, two of the transiting planets have been found to
  have companions: HAT-P-13b (whose companion is a planet;
  \citealt{Bakos_et_al_2009_2}), and HAT-P-7b (whose companion might
  be either a planet or low-mass star; \citealt{Pal_et_al_2008,
    Winn_et_al_2009_4}).  \citet{Batygin_et_al_2009_2} have coupled a
  three-body tidal orbital evolution model with the interior structure
  evolution of HAT-P-13b.  They found that a quasi-stationary solution
  of the type that we consider is viable in that system.
\item Furthermore, the systems we consider in this paper have all been
  detected fairly recently.  Companions could easily remain hidden if
  their orbital periods are sufficiently long compared with the
  temporal baseline for which the system has been observed, or if they
  induce sufficiently low reflex velocities in their host stars.
  Future observations and theoretical work could rule in or rule out
  the hypothesis that a companion excites the transiting planet's
  eccentricity.
\end{itemize}

\section{Theoretical Models for the Equilibrium Planetary Radius}
\label{sec:theo_Req}
If a gas giant planet has an additional power source heating its
interior, its radius will be larger than theory would otherwise
predict. The primary objective of this paper is to provide, for each
inflated planet listed in Section \ref{sec:properties}, the amount of
extra power that would be needed, in the steady state, to explain the
measured radii.\footnote{As \citet{Ibgui_and_Burrows_2009} show, if
  the steady-state assumption is dropped, earlier episodes of intense
  heating can result in a state such that the subsequent radius
  remains inflated despite lower interior heating at the present. This
  is due to thermal inertia; the convergence towards the non-inflated
  radius may be on hundreds of megayear timescales.}  If we assume
that the heating is due to tides, the energy dissipation rate within
the planet may be be represented as \citep{Peale_and_Cassen_1978,
  Murray_et_Dermott_1999, Bodenheimer_et_al_2001,
  Bodenheimer_et_al_2003, Gu_et_al_2004, Mardling_2007,
  Jackson_et_al_2008_3, Ibgui_and_Burrows_2009}:
\begin{equation}
\label{eq:Etides} 
\dot{E}_{\rm tide} = \phantom{-}  \left( \frac{63}{4} G^{3/2}M_{\ast}^{5/2} \right)  \frac{R_p^{5}}{a^{15/2}}  \frac{e^{2}}{Q'_p}\, ,
\end{equation}
where $G$ is the gravitational constant, $M_{\ast}$ is the mass of the
star, $R_p$ is the radius of the planet, $Q'_p$ is the tidal
dissipation factor \citep{Goldreich_1963}, and $e$ is the orbital
eccentricity. Note that the $e$-values of the systems considered here
(see Table~\ref{tab:transit_planets_data}) are small enough that this
equation, good to lowest order in eccentricity, is valid.  Henceforth,
unless otherwise stated, we will assume that the system has reached an
equilibrium state, such that the heating can be considered
constant. Maintaining tidal dissipation requires a mechanism to
sustain a nonzero eccentricity, as \citet{Bodenheimer_et_al_2001}
pointed out. A way to achieve approximately this scenario is via a
quasi-stationary orbit with $e$ and $a$ varying very slowly. If the
planet reaches its currently measured position in a timescale that is
short compared with the age of the system, typically in less than a
few Myr -- for example, through interactions with the protoplanetary
disk \citep{Lin_et_al_1996,Goldreich_and_Sari_2003} or by scattering
with other planets \citep{Ford_et_al_2003, Chatterjee_et_al_2008,
  Ford_and_Rasio_2008, Juric_and_Tremaine_2008} -- then its early
early evolution may be neglected. We also neglect any Kozai
interaction \citep{Wu_and_Murray_2003, Wu_2003,Wu_et_al_2007,
  Nagasawa_et_al_2008}. Once in its current orbit, its eccentricity
and semimajor axis may be maintained at quasi-constant values on Gyr
timescales through pumping by a third body (a planetary companion, for
instance; \citealt{Mardling_2007}). We note that the statistics of the
known exoplanet systems suggest that the Mardling mechanism is
plausible: about $10\%$ of the discovered exoplanetary systems are
known to be multiplanet systems
\citep{Wright_2009}\footnote{http://exoplanet.eu}.  Furthermore, at
least two transiting EGPs have companions: HAT-P-13b
\citep{Bakos_et_al_2009_2}, HAT-P-7b \citep{Pal_et_al_2008,
  Winn_et_al_2009_4}. Since the EGPs we consider are old compared with
the timescale for radius evolution (see
Table~\ref{tab:host_stars_data}), we treat the radius of each planet
as constant and equal to the equilibrium radius $R_{\rm eq}$ as
defined by \citet{Liu_et_al_2008}. If $e$, $a$, and $R_p$ are
(approximately) constant, then, per Equation~(\ref{eq:Etides}), tidal
heating is (approximately) constant as well.

\subsection{The Case of No Heavy-Element Central Core; the Effect of the Opacity}
\label{subsec:no_central_core}
We examine here the evolution of the equilibrium radius $R_{\rm eq}$
as a function of the extra-heating rate $\dot{E}_{\rm heating}$, while
excluding the presence of a heavy-element core ($M_{\rm core}=0$). We
compute the extra-heating rate required to fit the measured
radius. The results are summarized in Fig.~\ref{fig:fig1} for each of
the considered planets: TrES-4 (blue), WASP-4b (gray), WASP-6b (red),
WASP-12b (orange), and WASP-15b (green). HAT-P-12b does not appear
here because it requires a core, as we show in Section
\ref{subsec:central_core_solar}.  Table~\ref{tab:necessary_heating}
lists the extra-heating rate $\dot{E}_{\rm heating}$ needed to fit the
measured radii. It also lists the stellar irradiation\footnote{
  $\dot{E}_{\rm irradiation} = \pi {R_p}^2F_p$, where $F_p$ is the
  stellar irradiation flux.}  incident upon each planet in our study
(including HAT-P-12b). It compares $\dot{E}_{\rm heating}$ to
$\dot{E}_{\rm irradiation}$ by displaying the ratio $\dot{E}_{\rm
  heating}/\dot{E}_{\rm irradiation}$. Finally, it relates the orbital
eccentricity $e$ to the tidal dissipation factor $Q'_p$ through the
scaled ratio $(e/0.05)^2 / (Q'_p/10^5)$. We explore the effect of
enhanced atmospheric opacity on the required heating rate
$\dot{E}_{\rm heating}$, with our reference opacity corresponding to
solar metallicity abundance. As an example of enhanced atmospheric
opacity, we consider opacities corresponding to $10\times$solar
abundances, as described in \citet{burrows_et_al2007}.

The top left plot of the figure depicts $R_{\rm eq}$ as a function of
$\dot{E}_{\rm heating}$.  The thick horizontal segments are the best
measured values of the radii and the thick vertical segments are the
$1\sigma$ tolerances, as listed in
Table~\ref{tab:transit_planets_data}. The intersections between the
best measured values and the theoretical $R_{\rm eq}$ are represented
by filled circles. Two opacities are employed: solar (solid lines),
and $10\times$solar (dashed lines). As expected, the curves grow with
the heating rate: the greater the heating rate, the larger the
equilibrium radius. The bottom left parts of these curves are
horizontal. These correspond to the regimes where the heating effect
is negligible and, therefore, give the equilibrium radius without
extra-power. If we assume a solar opacity, we see that TrES-4 and
WASP-12b have the largest discrepancy between the standard model
radius without extra-power and the measured one (roughly the same
discrepancy for both planets). The difference is about $0.6~R_{J}$, or
$\sim$33$\%$ of the $\sim$$1.8~R_{J}$ measured radii. They require the
highest extra-heating rate $\dot{E}_{\rm heating} \sim 2\times
10^{-6}~L_{\sun}$ (see Table~\ref{tab:necessary_heating}) at
solar. WASP-6b is the least inflated planet or possibly a non-inflated
one. Its theoretical radius without extra heating is $1.20~R_{J}$,
very close to the $1.224^{+0.051}_{-0.052}~R_{J}$ measured radius (see
Table~\ref{tab:transit_planets_data}) and even within the 1$\sigma$
error. In between, WASP-4b and WASP-15b appear as ``moderately''
inflated with roughly the same relative gap of $13\%$ between the
measurement and the calculation at solar with no tides. Though the
relative gap is the same, the required heating at solar is a factor 16
greater for WASP-4b than for WASP-15b. This demonstrates that the
amount of heating to increase the radius depends not only on the
difference between the radius we want to reach and the radius without
heating, but also on the magnitude of the latter
\footnote{Extra heating for WASP-15b: $5.0\times10^{-9}~L_{\sun}$; for
  WASP-4b: $7.8\times10^{-8}~L_{\sun}$.}. The influence of planetary
opacity is indicated by the difference between the calculations at
solar (solid curves) and those at 10$\times$solar (dashed curves).
Models with higher atmospheric opacity require less heating, by
roughly a factor ten in these examples. This is because enhanced
opacity maintains the EGP's radius at higher values for longer times
\citep{burrows_et_al2007}. Note that if WASP-6b has 10$\times$solar
opacity, then it is ``not inflated'' (there is no filled circle). A
simulation (result not shown here) indicates that 3$\times$solar
opacity provides a good agreement of the radius within $1\sigma$.

The top right panel of Fig.~\ref{fig:fig1} plots the same ordinate as
the top left ($R_{\rm eq}$), but with the ratio $\dot{E}_{\rm
  heating}/\dot{E}_{\rm irradiation}$ as the x-axis. The main result
here is the small amount of extra heating that is required in
comparison with the irradiation rates; the ratios range from
$\sim$$10^{-6}$ to $\sim$$10^{-2}$.  A second issue is shown by the
curves depicting the cases of TrES-4 and WASP-12b. Both require
roughly the same $\dot{E}_{\rm heating}$
($\sim$$2\times10^{-6}~L_{\sun}$ at solar and $\sim$$10^{-7}~L_{\sun}$
at $10\times$solar), but the ratio $\dot{E}_{\rm heating}/\dot{E}_{\rm
  irradiation}$ is 3.5 higher for TrES-4 than for WASP-12b
($\dot{E}_{\rm irradiation}=1.2 \times 10^{-3} L_{\sun}$ for WASP-12b,
and $3.2\times 10^{-4} L_{\sun}$ for TrES-4).  In short, both planets
end up with the same equilibrium radius, while subjected to the same
tidal heating rate, whereas WASP-12b is 3.75 times more irradiated
than TrES-4. Everything else being equal, the more heavily irradiated
planet has the larger radius. Here, however, WASP-12b has a larger
mass than TrES-4 (the ratio is 1.5), and therefore is less affected by
the irradiation than TrES-4.

The top two panels of Fig.~\ref{fig:fig1} do not specify the origin of
the extra heating. If we assume that it is tidal heating, then using
Equation~(\ref{eq:Etides}) allows us to constrain the ratio
$e^2/Q'_p$. The bottom left side of Fig.~\ref{fig:fig1} plots the
equilibrium radius $R_{\rm eq}$ as a function of the scaled ratio
$(e/0.05)^2 / (Q'_p/10^5)$.  The curves have the same shapes as the
ones in the two previously discussed plots, especially the behavior at
the small values of this ratio that correspond to negligible tidal
heating, either because of very low eccentricity $e$ or because of
very large tidal dissipation factor $Q'_p$. This factor is a poorly
known parameter. We suppose that it has a constant value associated
with a given planet, though it depends on the tidal period and
therefore on the orbital period \citep{Ogilvie_and_Lin_2004,
  Wu_2005_1,Goodman_and_Lackner_2009,Spiegel_et_al_2010_2}. Note that
the stationary orbit hypothesis removes the orbital period dependence
of $Q'_p$, but it does not necessarily result in a constant value,
since $Q'_p$ also depends on other parameters such as $R_p$
\citep{Goodman_and_Lackner_2009}.

A precise measurement of $e$ and the use of these curves would help to
determine $Q'_p$ and, therefore, to constrain the models of tidal
dissipation in the structure of the giant planets
\citep{Ogilvie_and_Lin_2004, Wu_2005_1,
  Goodman_and_Lackner_2009}. Moreover, we provide in
Table~\ref{tab:limits_Qp} the rough estimates of possible values of
$Q'_p$ for the planets while considering the uncertainties in the
measured values of $e$ listed in Table~\ref{tab:transit_planets_data}.
For WASP-4b, WASP-6b, and WASP-12b, the $Q'_p$ values are of the order
of $10^{7}$ to $10^{9}$, higher than the $10^{5}$ to $10^{6}$ more
commonly assumed values, whether coming from Jupiter estimates
\citep{Goldreich+Soter_1966, Yoder_and_Peale_1981}, or from
theoretical arguments \citep{Ogilvie_and_Lin_2004}. On the other hand,
the values are very low for TrES-4 ($\sim 4 \times 10^{3}$ to $\sim 5
\times 10^{4}$). This suggests that either the planet is extremely
dissipative and will challenge theoretical models, or that it has a
much larger opacity than the $10\times$solar considered here (because,
as we have seen before, a larger opacity requires less heating and,
therefore, a higher $Q'_p$). Finally, for WASP-15b, the $Q'_p$ range
is quite large, from $\sim 10^{5}$ to $\sim 10^{9}$.  The high values
of $Q'_p$ suggest that all these planets, except perhaps TrES-4,
probably have a heavy-element core.  The presence of such a core
decreases the required values of $Q'_p$ to fit the measured radii.

The timescale necessary to reach an equilibrium radius has been
studied by \citet{Liu_et_al_2008}, based on generic sets of planets,
semimajor axes, and heating rates. Assuming that the planet receives
steady heating, their simulations show that the timescales can range
from as low as a few Myr for the most heavily irradiated planets to a
few Gyr for the less heavily irradiated ones.  In the bottom right
panel of Fig.~\ref{fig:fig1}, we address this issue for the particular
cases of the planets considered in this paper. This panel shows the
best fitting evolutionary curves, at solar and $10\times$solar
opacities, while maintaining the constant $\dot{E}_{\rm heating}$
assumption. The corresponding equilibrium radii and $\dot{E}_{\rm
  heating}$ are depicted by the thick dots in the three other
panels. TrES-4 and WASP-12b reach their equilibrium radius on a
timescale that is negligible compared to their estimated ages, even
negligible compared to 1 Gyr, at both opacities. Although the
equilibrium timescales for the three other planets at solar opacity,
WASP-4b, WASP-6b, and WASP-15, are longer, from 0.5 Gyr to a few Gyr,
we can reasonably assume that $R_{\rm eq}$ has been reached at their
putative ages. A larger opacity results in a longer timescale to reach
$R_{\rm eq}$, again because increased opacity slows down radius
shrinkage. As the case of WASP-15b at 10$\times$solar opacity shows,
the equilibrium radius $R_{\rm eq}$ might not be reached within the
age of the planet, if the heating remains constant and equal to the
power needed to match the observed radius within its $1\sigma$
tolerance.  One can imagine an initial episode of tidally coupled
radius-orbit evolution \citep{Ibgui_and_Burrows_2009,
  Ibgui_et_al_2009_2, Miller_et_al_2009} followed by a frozen orbital
position due to an interaction with a second planet.

\subsection{Effect of a Heavy-Element Central Core at Solar Opacity}
\label{subsec:central_core_solar}
In the previous subsection, we explored the effect of interior heating
on the equilibrium radius $R_{\rm eq}$, and we inferred the
relationship between $R_{\rm eq}$ and the ratio $e^2/Q'_p$.  In our
calculations so far, we have assumed that there is no heavy-element
central core in planets.  Now, we examine the influence such a core
would have on planetary radius and on the required steady-state
heating, at a given (solar) opacity. Figure~\ref{fig:fig2} summarizes
the results. The four panels are qualitatively similar to the ones in
Fig.~\ref{fig:fig1}. The difference is that instead of changing the
opacity, we change the core content from zero (no core) to an
arbitrary but sufficiently high core mass, such that its effect can be
appreciated. The solid curves are for the no-core and solar-opacity
case. They are exactly the same as in Fig.~\ref{fig:fig1}. The dotted
curves are for the case with cores, still at solar opacity. In
addition to the planets presented in Fig.~\ref{fig:fig1}, namely
TrES-4 (blue), WASP-4b (gray), WASP-6b (red), WASP-12b (orange), and
WASP-15b (green), we have added HAT-P-12b (cyan), which is not
inflated, but requires a core. The top left panel represents $R_{\rm
  eq}$ as a function of $\dot{E}_{\rm heating}$. Since the core acts
to shrink the radius, a higher heating rate is required to fit the
measured radius in the presence of a core.\footnote{Greater opacity,
  which also tends to increase the equilibrium radius, could partially
  substitute for a higher heating rate in the presence of a core.}
Note that, for HAT-P-12b, there is a large gap between the theoretical
radius without tidal heating ($\sim$$1.3~ R_{J}$) and the measured
radius ($\sim$$0.96~R_{J}$). As in Fig.~\ref{fig:fig1}, we also show
$R_{\rm eq}$ as a function of two different abscissae, $\dot{E}_{\rm
  heating}/\dot{E}_{\rm irradiation}$ in the top right panel, and the
scaled ratio $(e/0.05)^2 / (Q'_p/10^5)$ in the bottom left. The bottom
right panel shows the best fitting evolutionary curves, similar to the
corresponding panel in Fig.~\ref{fig:fig1}, but showing here the
effect of a core. This panel makes clear that the presence of a
heavy-element core shortens the timescale to reach the equilibrium
radius $R_{\rm eq}$. In fact, the larger the core mass, the higher the
value of the heating rate required to fit the radius, and the shorter
the timescale to reach equilibrium.

\subsection{Models with Compatible Core Mass - Heating Rate Pairs at solar and $10\times$solar}
\label{subsec:compatible_models}
In the calculations so far (Section \ref{subsec:no_central_core} and
Section \ref{subsec:central_core_solar}), we have mainly examined the
qualitative effects of atmospheric opacity and of a heavy-element core
on the extra interior heating required for steady-state models to
match the observed planetary radii.  The upshot is that the
theoretical radius is increased by a higher heating rate $\dot{E}_{\rm
  heating}$, and by greater atmospheric opacity, but reduced by a
larger core mass $M_{\rm core}$. Therefore, for each planet, for a
given atmospheric opacity, we can find a range of possible couples
$\{M_{\rm core},\dot{E}_{\rm heating}\}$ that allow our model to fit
the planetary radius. The objective of this section is to explore
these possibilities and to translate this degeneracy in terms of
possible couples $\{M_{\rm core},e^2/Q'_p\}$, at solar or enhanced
opacity.  Note that our model decouples the opacity of the planetary
atmosphere from the core mass. This is an approximation. Since planets
originate from the same protostellar, protoplanetary disk as their
stars, it is reasonable to assume that there is a correlation between
host star metallicity and bulk planetary metallicity -- though, as the
different bulk metallicities of the planets in our solar system
attest, any such correlation involves nonzero scatter. Furthermore,
though there is naturally a correlation between atmospheric
metallicity and atmospheric opacity, the details of a given planet's
atmospheric chemistry and thermophysics influence how its cocktail of
elements translates into radiative opacity.  It is important to note
that each of these correlations is uncertain. In the same way as in
Section \ref{subsec:no_central_core}, our reference case is solar
opacity and we consider $10\times$solar as an example of enhanced
atmospheric opacity. Figure~\ref{fig:fig3}, the central figure of this
paper, and both Tables~\ref{tab:limits_Mcore_HAT-P-12b_TrES_4_WASP-4b}
and \ref{tab:limits_Mcore_WASP-6b_WASP-12b_WASP-15b}, summarize the
results.

The top left panel of Fig.~\ref{fig:fig3} shows the required heating
$\dot{E}_{\rm heating}$ as a function of the core mass $M_{\rm core}$,
for HAT-P-12b (cyan), TrES-4 (blue), WASP-4b (gray), WASP-6b (red),
WASP-12b (orange), and WASP-15b (green), at two atmospheric opacities
-- solar (solid), and $10\times$solar (dashed). As expected,
$\dot{E}_{\rm heating}$ increases with $M_{\rm core}$, but at rates
that differ from one planet to another. HAT-P-12b appears to be the
most sensitive with the highest relative slope $d\ln\dot{E}_{\rm
  heating} / dM_{\rm core} \sim 0.11~M_{\earth}^{-1}$.  This could be
explained by the fact that it is by far the lightest planet of the
list, with $M_{p} \approx 0.21~M_{J}$
(cf. Table~\ref{tab:transit_planets_data}), and thus the most
sensitive to an extra-heating power. Moreover, the figure shows that
it is the only clearly non-inflated planet, since even without heating
a significant core mass is necessary to fit the radius
($31~M_{\earth}$, which is 46$\%$ of the total mass of the planet; see
Table~\ref{tab:limits_Mcore_HAT-P-12b_TrES_4_WASP-4b}).  On the other
hand, WASP-12b, the most massive of the planets of the list ($1.41
M_{J}$), is the least sensitive to an extra-heating power, with
$d\ln\dot{E}_{\rm heating} / dM_{\rm core} \sim
0.010~M_{\earth}^{-1}$, which is ten times smaller than the relative
slope for HAT-P-12b. An enhancement of the opacity, for a given
$M_{\rm core}$, results in the decrease of the required $\dot{E}_{\rm
  heating}$, as already noted in section \ref{subsec:no_central_core}
for a planet without a core. WASP-6b at $10\times$solar would seem a
non-inflated planet, since it would then require a small core even
without heating ($M_{\rm core}=6~M_{\earth}$). The WASP-15b radius can
be fit without heating and without a core, if its opacity is equal to
10$\times$solar. Note also that at this opacity and without a core,
the equilibrium radius $R_{\rm eq}$ cannot be reached as shown by
Fig.~\ref{fig:fig3} in Section \ref{subsec:no_central_core}.
Therefore, if WASP-15b has supersolar opacity and a small core (or
none), then it has probably not been experiencing steady-state
interior heating since the early epochs after its birth. It might,
nonetheless, be undergoing quasi steady-state interior heating at the
present.  Another point emphasized by this panel is the wide range of
core masses compatible with the measured radius. For example, we ran
simulations for WASP-12b with a core mass up to 300~$M_{\earth}$
(67$\%$ of the total planet's mass) at solar, and up to 351
$M_{\earth}$ (78$\%$ of the total mass) at $10\times$solar. These
masses are quite large, which raises the following question: Is there
a theoretical upper limit for the core masses, compatible with the
models of planetary formation?  \citet{Pont_et_al_2009_1}, who propose
some radius evolutionary models for HD 80606b, state that core masses
above 200~$M_{\earth}$ might be unrealistic, given the constraints of
planetary formation models \citep{Ikoma_et_al_2006}. In particular,
they point out a competition between scattering and accretion of
planetesimals \citep{Guillot_and_Gladman_2000}. In the future, such
models might provide upper limits on $M_{\rm core}$, which, in
combination with our calculations presented in Fig.~\ref{fig:fig3} and
Tables~\ref{tab:limits_Mcore_HAT-P-12b_TrES_4_WASP-4b} and
\ref{tab:limits_Mcore_WASP-6b_WASP-12b_WASP-15b}, may help to decipher
the structures of transiting EGPs.

The top right panel of Fig.~\ref{fig:fig3} shows the ratio
$\dot{E}_{\rm heating}/\dot{E}_{\rm irradiation}$ as a function of
$M_{\rm core}$.  This ratio generally remains small, from $10^{-5}$ to
$10^{-1}$, as already noticed with the models without core (Section
\ref{subsec:no_central_core}).

The link between $M_{\rm core}$ and the poorly known parameters $e$
and $Q'_p$, through the ratio $e^2/Q'_p$, are shown in the bottom left
panel of Fig.~\ref{fig:fig3}. Note that we plot the scaled ratio
$(e/0.05)^2 / (Q'_p/10^5)$ . This relation (see also tabulated values
in Tables \ref{tab:limits_Mcore_HAT-P-12b_TrES_4_WASP-4b} and
\ref{tab:limits_Mcore_WASP-6b_WASP-12b_WASP-15b}) could be used to
constrain the models of tidal dissipation in planets in conjunction
with the planetary structure models. Indeed, a precise enough
measurement of the orbital eccentricity $e$ of these planets would
result in a direct link between $Q'_p$ and $M_{\rm core}$. We provide
such a link, with the best current measured estimations of $e$ in the
bottom right panel of Fig.~\ref{fig:fig3}. It is important to bear
this in mind that uncertainties in $e$ translate into uncertainties in
these curves.  For a given planet in the stationary state, the ratio
$e^2/Q'_p$ is a constant. We can therefore link the uncertainties
$\Delta e$ and $\Delta Q'_p$, by writing $e^2/Q'_p=(e+\Delta
e)^2/(Q'_p+\Delta Q'_p)$, which yields:
\begin{equation}
\label{eq:d_e_Q}
\frac{\Delta Q'_p}{Q'_p} = \frac{\Delta e}{e}  \left( 2+ \frac{\Delta e}{e} \right) \, .
\end{equation}
Note that such a formulation acknowledges that the relative
uncertainties, $\Delta e/e$ and $\Delta Q'_p/Q'_p$, can be large.
Lower and upper limits of the measured $e$ are available for WASP-6b,
WASP-12b, WASP-15b (Table~\ref{tab:transit_planets_data}). We have
only upper limits for the other planets. We infer lower and upper
limits on $\Delta \log Q'_p=\log(1+\Delta Q'_p /Q'_p)$:
($-0.35,+0.25$) for WASP-6b, ($-0.31,+0.23$) for WASP-12b, and
($-0.71,+0.38$) for WASP-15b.  Given the low precision of $e$, these
curves have to be considered with caution. However, they provide some
useful information. In particular, the $Q'_p$s that allow the radii to
be fit fall within a wide range, from $10^{3}$ to $10^{9.5}$, and
appear to be a property that differs from one planet to
another. Moreover, for a given $M_{\rm core}$, the larger the
atmospheric opacity, the larger the $Q'_p$. This is unsurprising
since, as we have previously seen, an enhanced opacity results in a
larger radius. Therefore, less tidal heating is required (hence the
larger $Q'_p$). TrES-4 is the most problematic planet. The compatible
$Q'_p$s at solar opacity are too low, $10^{3}$ to $10^{3.5}$ for
$e=0.01$ at solar opacity. \citet{Knutson_et_al_2009_1} derive a $3
\sigma$ upper limit $|e\cos(\omega)|$ of ~0.0058, so $e$ is likely to
be lower than 0.01, therefore reducing $Q'_p$ further (in order to
maintain the same ratio $e^2/Q'_p$). Estimates for Jupiter are around
$10^{5}$ to $10^{6}$ \citep{Goldreich+Soter_1966,
  Yoder_and_Peale_1981}, and \citet{Ogilvie_and_Lin_2004} provide
theoretical motivations for $Q'_p$ values around $10^{5}$. The
assumption of $10\times$solar atmospheric opacity increases roughly by
a factor 10 the value of $Q'_p$. Perhaps TrES-4 has a higher opacity.
Another possibility is that we have misinterpreted the radius of this
planet. It might have rings, although icy rings would probably
sublimate because of the intense irradiation, and dusty rings would
probably plunge quickly into the star under the influence of the
Poynting-Robertson drag \citep{poynting1903, robertson1937}. It may
also be conceivable that its star is smaller than has been inferred,
which, if so, would imply that the planet is correspondingly smaller
than has been inferred.  If we assume that values of $Q'_p$ span the
$10^{5}$-$10^{6}$ range, we can infer possible core masses.  We have
for WASP-6b an $M_{\rm core}$ between roughly 50 and 90 $M_{\earth}$
if the opacity is solar, and between 80 and 110 $M_{\earth}$ if it is
$10\times$solar. For WASP-15b, we have $M_{\rm core}$ between roughly
30 and 70 $M_{\earth}$ at solar, and between 70 and 115 $M_{\earth}$
at $10\times$solar.  WASP-12b requires $Q'_p$ values larger than
$10^{6}$. Given the lower limit $\Delta \log Q'_p=-0.71$ (see above),
$Q'_p$ values of the order or smaller than $10^{6}$ might be reached,
but for a quite large $M_{\rm core} (\gtrsim 150 M_{\earth})$ at
solar.  Note that it could seem surprising that WASP-12b and TrES-4,
which have very similar radii and comparable masses
(Table~\ref{tab:transit_planets_data}), fit with so different $Q'_p$
parameters: the order of magnitude of $Q'_p$ for WASP-12b is roughly
four times greater than the one for TrES-4
(Table~\ref{tab:limits_Mcore_HAT-P-12b_TrES_4_WASP-4b}
and~\ref{tab:limits_Mcore_WASP-6b_WASP-12b_WASP-15b}, and
Figure~\ref{fig:fig3}). This can be explained by
Equation~(\ref{eq:Etides}) and by the fact that the heating rates,
$\dot{E}_{\rm heating}$, necessary to fit the observed radii, are
comparable for these two planets
(cf. Sections~\ref{subsec:no_central_core}
and~\ref{subsec:central_core_solar}, and upper left panels of
Figures~\ref{fig:fig1} and~\ref{fig:fig2}). With comparable host star
masses (Table~\ref{tab:host_stars_data}) and planetary radii, the
ratio of the $Q'_p$ is given by $[(e^2/a^{7.5})_{\rm WASP-12b} /
  (e^2/a^{7.5})_{\rm TrES-4}]$, which is roughly $10^{3.6}$, assuming
for $a$ and $e$ the values in Table~\ref{tab:transit_planets_data}.
Note that current theories on the mechanisms of tidal dissipation in
EGPs suggest that the $Q'_p$ parameter depends sensitively on the
structure and atmosphere of the planet and that it could vary
significantly from planet to planet, even though they might otherwise
seem structurally similar, as we might think for WASP-12b and TrES-4.
\citet{Goodman_and_Lackner_2009} argue that $Q'_p$ might vary
significantly depending on the possible presence of a solid core and
its size, completely unconstrained for these two planets.
Furthermore, \citet{Ogilvie_and_Lin_2004} and \citet{Wu_2005_1} argue
that $Q'_p$ could depend quite sensitively on the tidal forcing
frequency.  Since these two planets have very different orbital
periods ($P_{\rm TrES-4} \sim 3 \times P_{\rm WASP-12b}$; see
Table~\ref{tab:transit_planets_data}) there is no reason to expect the
Q's for these two planets to be similar; orders of magnitude
ambiguity, even for planets of comparable mass and radius, are
entirely possible.  In addition, the host stars' metallicities are
different (around 0.30~dex for WASP-12b and 0.14~dex for TrES-4).  It
is not inconceivable that the atmospheric opacities of the two planets
are different from solar (and from each other), which would change the
values of $Q'_p$ that are required, in our model, in order for tidal
heating to explain the inflated radii.  Finally, the inferred values
of $Q'_p$ for WASP-4b, and even more for HAT-P-12b, might be larger
than is realistic.  These results are based only on upper limits for
$e$, which is too poor a constraint.  Lower values of $e$ reduce the
values of $Q'_p$ that are required for our model to explain the radii.
We note that there is a statistical bias toward overestimating
eccentricities from radial velocity measurements
\citep{Shen_and_Turner_2008}.  Substantially improved measurements of
orbital eccentricity, together with improved knowledge of atmospheric
opacity, would allow us to constrain $Q'_p$, if the steady-state model
is appropriate. As an order-of-magnitude estimate,
Equation~(\ref{eq:d_e_Q}) shows that an error $\Delta \log Q'_p
\approx \pm 0.25$ could be achieved with a relative error on $e$ of $
\approx \pm 33\%$. Planetary data, in
Table~\ref{tab:transit_planets_data}, show that this constraint is
satisfied by the observed eccentricities of WASP-6b and WASP-12b only.
We stress that tests of the steady-state scenario must be done on a
planet-by-planet basis.

\section{Conclusions and Discussion}
\label{sec:conclusion}
We have proposed in this paper a stationary tidal heating scenario to
explain the radii of the inflated transiting EGPs WASP-4b, WASP-6b,
WASP-12b, WASP-15b, and TrES-4. A constant heating rate may be
achieved with quasi-constant eccentricity and semimajor axis, and
after the planetary radius has reached its equilibrium value. Such a
scenario might explain the inflated radii of some transiting EGPs. We
have calculated the amount of additional interior heating that is
required to fit the radius of each planet, and explored how it depends
on the planet's atmospheric opacity and on a heavy-element central
core. There is a degeneracy between the required heating rate to fit
the radius, the atmospheric opacity, and the heavy-element core mass
inside the planet. In terms of tidal heating, there is, for a given
opacity, a locus of points $\{M_{\rm core},e^2/Q'_p\}$ that can lead
to the same radius. For this reason, a substantial improvement in the
precision of measured orbital eccentricities would transform this
degeneracy into a reliable $\{M_{\rm core},Q'_p\}$
degeneracy. Combined with theories of tidal dissipation inside the
planets that constrain the values of $Q'_p$, and with planet formation
models that may also constrain the values of $M_{\rm core}$, this
scenario could represent an explanation of these inflated radii based
on tidal heating.

Our general conclusions are the following:
\begin{itemize}\itemsep0cm
\item The heating rate, $\dot{E}_{\rm heating}$, necessary to fit an
  observed inflated radius is generally small in comparison to the
  irradiation rate, $\dot{E}_{\rm irradiation}$, with a ratio
  $\dot{E}_{\rm heating}/\dot{E}_{\rm irradiation}$ ranging from
  $10^{-6}$ to $10^{-1}$.
\item The higher the heating rate, the larger the equilibrium
  radius. On the other hand, the larger the core mass, the smaller the
  equilibrium radius, and the higher the required heating rate in
  order to fit the measured radius. In other words, $\dot{E}_{\rm
    heating}$ increases with $M_{\rm core}$.
\item Tidal heating might provide the extra interior power that
  inflates some planets.  If tides are the extra power source, then
  the tidal dissipation factor $Q'_p$ is a decreasing function of core
  mass $M_{\rm core}$, but an increasing function of the atmospheric
  opacity. It may also be different for different planets.
\item The more massive the planet, the less its radius is sensitive to
  extra heating.
\item If the steady-state scenario presented in this paper obtains in
  a planetary system, the presence and the characterization of a
  central heavy-element core in that planet can be constrained by a
  better knowledge of the eccentricity, the planet's atmospheric
  opacity, and the tidal dissipation in its interior governed by
  $Q'_p$.
\item It would be interesting to revisit the work of
  \citet{Mardling_2007} in the specific instances of the planets under
  consideration in this paper, so as to characterize what possible
  companions might cause quasi-constant eccentricity and semimajor
  axis. Better constraints from radial velocity data will reveal which
  such companions, if any, might be present.
\end{itemize}

The applicability of the steady-state scenario should be examined for
each planet individually. Assuming the current best measured estimates
of the eccentricities, and being fully aware of the poor knowledge of
this parameter and, therefore, that the following comments will need
to be revised when more accurate measurements are available, we can
state more specifically, planet by planet, that:
\begin{itemize}\itemsep0cm
\item TrES-4, for which we assume $e$=0.01, is the most difficult to
  reconcile with the steady-state scenario. The tidal dissipation
  parameter $Q'_p$ might be $10^{3.5}$ for a solar atmospheric
  opacity, lower than theoretical models or empirical determinations
  would generally suggest.  If, as is likely, the actual eccentricity
  is lower than 0.01, the required value of $Q'_p$, in the context of
  our model, would be even lower.  On the other hand, if the
  atmospheric opacity is greater than solar, the planet would cool
  more slowly and the required value of $Q'_p$, in our model, would be
  greater (for instance, $10^{4.7}$ for a $10\times$solar opacity).
  For the purpose of constraining evolutionary models of this planet,
  it would be useful to study in greater depth both its eccentricity,
  which is not well constrained, and the composition and opacity of
  its atmosphere.
\item WASP-4b can be fit, but with values of $Q'_p$ or $M_{\rm core}$
  that are higher than we would expect. Its actual eccentricity might
  be noticeably lower than the upper limit currently available.
\item WASP-6b can be fit within $1\sigma$ with a $3\times$solar
  opacity, without any extra heating. Consequently, a lower opacity
  involves an extra heating.
\item WASP-12b requires either high values of $Q'_p$ or extremely high
  core masses, such as 300~$M_{\earth}$. However, such a high $M_{\rm
    core}$ might be ruled out by planetary formation models.
\item WASP-15b can be fit with $Q'_p$ ranging in the expected interval
  $10^{5}-10^{6}$ and with core masses below $\sim$$100 M_{\earth}$.
\item HAT-P-12b, the only non-inflated planet, can be fit without
  invoking tidal heating, and with a core mass smaller than $\sim$$50
  M_{\earth}$. However, models involving tidal heating and a central
  core are also compatible with its observed properties, within the
  steady-state assumption of this paper.
\end{itemize}

There are several caveats that the reader should bear in mind.
Details of tidal dissipation theory remain to be determined,
specifically the mechanism and location of heat deposition (in the
radiative mantle or in the convective envelope).  Also, we need to
better understand the difference between day and night cooling.  This
may require full three dimensional general circulation models with
radiative transfer.  Finally, we have assumed that the core mass and
the atmospheric opacity are independent.  Broadly speaking, it seems
likely that these properties are somewhat connected, due to a probable
correlation of the compositions of the central core, the convective
envelope, and the atmosphere. However, the lack of a unique mapping
between core mass and atmospheric opacity might limit the extent to
which future investigations might be able to treat these variables as
coupled.

The fact that some transiting extrasolar giant planets have inflated
radii continues to pose a theoretical challenge.  The stationary tidal
heating scenario described in this paper might explain the puzzle in
some cases.  Further observational constraints will clarify the
viability of this scenario.  To conclude, we mention the possibility
of an alternative model, that might also be able to explain the
inflated radii of some of the planets under consideration in this
paper -- namely, the coupled evolution of the planetary radius and its
orbit, in the case of a two-body gravitational and tidal interaction
\citep{Ibgui_and_Burrows_2009, Ibgui_et_al_2009_2,
  Miller_et_al_2009}. At this stage, both scenarios appear to be
plausible.

\acknowledgements

We thank Ivan Hubeny for help on issues concerning the computing of
the atmospheric models for the boundary conditions.  We thank Jeremy
Goodman, Rosemary Mardling, and Gordon Ogilvie for their instructive
insights into the physical modeling of the tidal dissipation factors.
We also thank Jason Nordhaus for useful discussions.  We thank the
anonymous referee for constructive comments and suggestions that
improved the manuscript.  The authors are pleased to acknowledge that
part of the work reported for this paper was performed at the TIGRESS
high performance computer center at Princeton University, which is
jointly supported by the Princeton Institute for Computational Science
and Engineering and the Princeton University Office of Information
Technology.  This study was supported by NASA grant NNX07AG80G and
under JPL/Spitzer Agreements 1328092, 1348668, and 1312647.

\bibliography{biblio}

\begin{thebibliography}{100}
\expandafter\ifx\csname natexlab\endcsname\relax\def\natexlab#1{#1}\fi

\bibitem[{{Alonso} {et~al.}(2008){Alonso}, {Auvergne}, {Baglin}, {Ollivier},
  {Moutou}, {Rouan}, {Deeg}, {Aigrain}, {Almenara}, {Barbieri}, {Barge},
  {Benz}, {Bord{\'e}}, {Bouchy}, {de La Reza}, {Deleuil}, {Dvorak}, {Erikson},
  {Fridlund}, {Gillon}, {Gondoin}, {Guillot}, {Hatzes}, {H{\'e}brard},
  {Kabath}, {Jorda}, {Lammer}, {L{\'e}ger}, {Llebaria}, {Loeillet}, {Magain},
  {Mayor}, {Mazeh}, {P{\"a}tzold}, {Pepe}, {Pont}, {Queloz}, {Rauer},
  {Shporer}, {Schneider}, {Stecklum}, {Udry}, \&
  {Wuchterl}}]{Alonso_et_al_2008}
{Alonso}, R., {Auvergne}, M., {Baglin}, A., {Ollivier}, M., {Moutou}, C.,
  {Rouan}, D., {Deeg}, H.~J., {Aigrain}, S., {Almenara}, J.~M., {Barbieri}, M.,
  {Barge}, P., {Benz}, W., {Bord{\'e}}, P., {Bouchy}, F., {de La Reza}, R.,
  {Deleuil}, M., {Dvorak}, R., {Erikson}, A., {Fridlund}, M., {Gillon}, M.,
  {Gondoin}, P., {Guillot}, T., {Hatzes}, A., {H{\'e}brard}, G., {Kabath}, P.,
  {Jorda}, L., {Lammer}, H., {L{\'e}ger}, A., {Llebaria}, A., {Loeillet}, B.,
  {Magain}, P., {Mayor}, M., {Mazeh}, T., {P{\"a}tzold}, M., {Pepe}, F.,
  {Pont}, F., {Queloz}, D., {Rauer}, H., {Shporer}, A., {Schneider}, J.,
  {Stecklum}, B., {Udry}, S., \& {Wuchterl}, G. 2008, \aap, 482, L21

\bibitem[{{Anderson} {et~al.}(2010){Anderson}, {Hellier}, {Gillon}, {Triaud},
  {Smalley}, {Hebb}, {Collier Cameron}, {Maxted}, {Queloz}, {West}, {Bentley},
  {Enoch}, {Horne}, {Lister}, {Mayor}, {Parley}, {Pepe}, {Pollacco},
  {S{\'e}gransan}, {Udry}, \& {Wilson}}]{Anderson_et_al_2009}
{Anderson}, D.~R., {Hellier}, C., {Gillon}, M., {Triaud}, A.~H.~M.~J.,
  {Smalley}, B., {Hebb}, L., {Collier Cameron}, A., {Maxted}, P.~F.~L.,
  {Queloz}, D., {West}, R.~G., {Bentley}, S.~J., {Enoch}, B., {Horne}, K.,
  {Lister}, T.~A., {Mayor}, M., {Parley}, N.~R., {Pepe}, F., {Pollacco}, D.,
  {S{\'e}gransan}, D., {Udry}, S., \& {Wilson}, D.~M. 2010, \apj, 709, 159

\bibitem[{{Bakos} {et~al.}(2009){Bakos}, {Howard}, {Noyes}, {Hartman},
  {Torres}, {Kov{\'a}cs}, {Fischer}, {Latham}, {Johnson}, {Marcy}, {Sasselov},
  {Stefanik}, {Sip{\H o}cz}, {Kov{\'a}cs}, {Esquerdo}, {P{\'a}l},
  {L{\'a}z{\'a}r}, {Papp}, \& {S{\'a}ri}}]{Bakos_et_al_2009_2}
{Bakos}, G.~{\'A}., {Howard}, A.~W., {Noyes}, R.~W., {Hartman}, J., {Torres},
  G., {Kov{\'a}cs}, G., {Fischer}, D.~A., {Latham}, D.~W., {Johnson}, J.~A.,
  {Marcy}, G.~W., {Sasselov}, D.~D., {Stefanik}, R.~P., {Sip{\H o}cz}, B.,
  {Kov{\'a}cs}, G., {Esquerdo}, G.~A., {P{\'a}l}, A., {L{\'a}z{\'a}r}, J.,
  {Papp}, I., \& {S{\'a}ri}, P. 2009, \apj, 707, 446

\bibitem[{{Bakos} {et~al.}(2007){Bakos}, {Noyes}, {Kov{\'a}cs}, {Latham},
  {Sasselov}, {Torres}, {Fischer}, {Stefanik}, {Sato}, {Johnson}, {P{\'a}l},
  {Marcy}, {Butler}, {Esquerdo}, {Stanek}, {L{\'a}z{\'a}r}, {Papp}, {S{\'a}ri},
  \& {Sip{\H o}cz}}]{bakos_et_al2007a}
{Bakos}, G.~{\'A}., {Noyes}, R.~W., {Kov{\'a}cs}, G., {Latham}, D.~W.,
  {Sasselov}, D.~D., {Torres}, G., {Fischer}, D.~A., {Stefanik}, R.~P., {Sato},
  B., {Johnson}, J.~A., {P{\'a}l}, A., {Marcy}, G.~W., {Butler}, R.~P.,
  {Esquerdo}, G.~A., {Stanek}, K.~Z., {L{\'a}z{\'a}r}, J., {Papp}, I.,
  {S{\'a}ri}, P., \& {Sip{\H o}cz}, B. 2007, \apj, 656, 552

\bibitem[{{Baraffe} {et~al.}(2006){Baraffe}, {Alibert}, {Chabrier}, \&
  {Benz}}]{Baraffe_et_al_2006}
{Baraffe}, I., {Alibert}, Y., {Chabrier}, G., \& {Benz}, W. 2006, \aap, 450,
  1221

\bibitem[{{Baraffe} {et~al.}(2008){Baraffe}, {Chabrier}, \&
  {Barman}}]{Baraffe_et_al_2008}
{Baraffe}, I., {Chabrier}, G., \& {Barman}, T. 2008, \aap, 482, 315

\bibitem[{{Baraffe} {et~al.}(2003){Baraffe}, {Chabrier}, {Barman}, {Allard}, \&
  {Hauschildt}}]{Baraffe_et_al_2003}
{Baraffe}, I., {Chabrier}, G., {Barman}, T.~S., {Allard}, F., \& {Hauschildt},
  P.~H. 2003, \aap, 402, 701

\bibitem[{{Baraffe} {et~al.}(2005){Baraffe}, {Chabrier}, {Barman}, {Selsis},
  {Allard}, \& {Hauschildt}}]{Baraffe_et_al_2005}
{Baraffe}, I., {Chabrier}, G., {Barman}, T.~S., {Selsis}, F., {Allard}, F., \&
  {Hauschildt}, P.~H. 2005, \aap, 436, L47

\bibitem[{{Baraffe} {et~al.}(2004){Baraffe}, {Selsis}, {Chabrier}, {Barman},
  {Allard}, {Hauschildt}, \& {Lammer}}]{Baraffe_et_al_2004}
{Baraffe}, I., {Selsis}, F., {Chabrier}, G., {Barman}, T.~S., {Allard}, F.,
  {Hauschildt}, P.~H., \& {Lammer}, H. 2004, \aap, 419, L13

\bibitem[{{Barge} {et~al.}(2008){Barge}, {Baglin}, {Auvergne}, {Rauer},
  {L{\'e}ger}, {Schneider}, {Pont}, {Aigrain}, {Almenara}, {Alonso},
  {Barbieri}, {Bord{\'e}}, {Bouchy}, {Deeg}, {La Reza}, {Deleuil}, {Dvorak},
  {Erikson}, {Fridlund}, {Gillon}, {Gondoin}, {Guillot}, {Hatzes}, {Hebrard},
  {Jorda}, {Kabath}, {Lammer}, {Llebaria}, {Loeillet}, {Magain}, {Mazeh},
  {Moutou}, {Ollivier}, {P{\"a}tzold}, {Queloz}, {Rouan}, {Shporer}, \&
  {Wuchterl}}]{Barge_et_al_2008}
{Barge}, P., {Baglin}, A., {Auvergne}, M., {Rauer}, H., {L{\'e}ger}, A.,
  {Schneider}, J., {Pont}, F., {Aigrain}, S., {Almenara}, J.-M., {Alonso}, R.,
  {Barbieri}, M., {Bord{\'e}}, P., {Bouchy}, F., {Deeg}, H.~J., {La Reza}, D.,
  {Deleuil}, M., {Dvorak}, R., {Erikson}, A., {Fridlund}, M., {Gillon}, M.,
  {Gondoin}, P., {Guillot}, T., {Hatzes}, A., {Hebrard}, G., {Jorda}, L.,
  {Kabath}, P., {Lammer}, H., {Llebaria}, A., {Loeillet}, B., {Magain}, P.,
  {Mazeh}, T., {Moutou}, C., {Ollivier}, M., {P{\"a}tzold}, M., {Queloz}, D.,
  {Rouan}, D., {Shporer}, A., \& {Wuchterl}, G. 2008, \aap, 482, L17

\bibitem[{{Batygin} {et~al.}(2009){Batygin}, {Bodenheimer}, \&
  {Laughlin}}]{Batygin_et_al_2009_2}
{Batygin}, K., {Bodenheimer}, P., \& {Laughlin}, G. 2009, \apjl, 704, L49

\bibitem[{{Bodenheimer} {et~al.}(2003){Bodenheimer}, {Laughlin}, \&
  {Lin}}]{Bodenheimer_et_al_2003}
{Bodenheimer}, P., {Laughlin}, G., \& {Lin}, D.~N.~C. 2003, \apj, 592, 555

\bibitem[{{Bodenheimer} {et~al.}(2001){Bodenheimer}, {Lin}, \&
  {Mardling}}]{Bodenheimer_et_al_2001}
{Bodenheimer}, P., {Lin}, D.~N.~C., \& {Mardling}, R.~A. 2001, \apj, 548, 466

\bibitem[{{Burrows} {et~al.}(2008{\natexlab{a}}){Burrows}, {Budaj}, \&
  {Hubeny}}]{Burrows_et_al_2008_1}
{Burrows}, A., {Budaj}, J., \& {Hubeny}, I. 2008{\natexlab{a}}, \apj, 678, 1436

\bibitem[{{Burrows} {et~al.}(2000){Burrows}, {Guillot}, {Hubbard}, {Marley},
  {Saumon}, {Lunine}, \& {Sudarsky}}]{burrows_et_al2000}
{Burrows}, A., {Guillot}, T., {Hubbard}, W.~B., {Marley}, M.~S., {Saumon}, D.,
  {Lunine}, J.~I., \& {Sudarsky}, D. 2000, \apjl, 534, L97

\bibitem[{{Burrows} {et~al.}(1993){Burrows}, {Hubbard}, {Saumon}, \&
  {Lunine}}]{Burrows_et_al_1993}
{Burrows}, A., {Hubbard}, W.~B., {Saumon}, D., \& {Lunine}, J.~I. 1993, \apj,
  406, 158

\bibitem[{{Burrows} {et~al.}(2007){Burrows}, {Hubeny}, {Budaj}, \&
  {Hubbard}}]{burrows_et_al2007}
{Burrows}, A., {Hubeny}, I., {Budaj}, J., \& {Hubbard}, W.~B. 2007, \apj, 661,
  502

\bibitem[{{Burrows} {et~al.}(2004){Burrows}, {Hubeny}, {Hubbard}, {Sudarsky},
  \& {Fortney}}]{burrows_et_al2004}
{Burrows}, A., {Hubeny}, I., {Hubbard}, W.~B., {Sudarsky}, D., \& {Fortney},
  J.~J. 2004, \apjl, 610, L53

\bibitem[{{Burrows} {et~al.}(2008{\natexlab{b}}){Burrows}, {Ibgui}, \&
  {Hubeny}}]{Burrows_et_al_2008_2}
{Burrows}, A., {Ibgui}, L., \& {Hubeny}, I. 2008{\natexlab{b}}, \apj, 682, 1277

\bibitem[{{Burrows} {et~al.}(1997){Burrows}, {Marley}, {Hubbard}, {Lunine},
  {Guillot}, {Saumon}, {Freedman}, {Sudarsky}, \& {Sharp}}]{Burrows_et_al_1997}
{Burrows}, A., {Marley}, M., {Hubbard}, W.~B., {Lunine}, J.~I., {Guillot}, T.,
  {Saumon}, D., {Freedman}, R., {Sudarsky}, D., \& {Sharp}, C. 1997, \apj, 491,
  856

\bibitem[{{Burrows} {et~al.}(2003){Burrows}, {Sudarsky}, \&
  {Hubbard}}]{burrows_et_al2003}
{Burrows}, A., {Sudarsky}, D., \& {Hubbard}, W.~B. 2003, \apj, 594, 545

\bibitem[{{Chabrier} \& {Baraffe}(2007)}]{chabrier+baraffe2007}
{Chabrier}, G., \& {Baraffe}, I. 2007, \apjl, 661, L81

\bibitem[{{Chabrier} {et~al.}(2004){Chabrier}, {Barman}, {Baraffe}, {Allard},
  \& {Hauschildt}}]{Chabrier_et_al_2004}
{Chabrier}, G., {Barman}, T., {Baraffe}, I., {Allard}, F., \& {Hauschildt},
  P.~H. 2004, \apjl, 603, L53

\bibitem[{{Charbonneau} {et~al.}(2000){Charbonneau}, {Brown}, {Latham}, \&
  {Mayor}}]{charbonneau_et_al2000}
{Charbonneau}, D., {Brown}, T.~M., {Latham}, D.~W., \& {Mayor}, M. 2000, \apjl,
  529, L45

\bibitem[{{Chatterjee} {et~al.}(2008){Chatterjee}, {Ford}, {Matsumura}, \&
  {Rasio}}]{Chatterjee_et_al_2008}
{Chatterjee}, S., {Ford}, E.~B., {Matsumura}, S., \& {Rasio}, F.~A. 2008, \apj,
  686, 580

\bibitem[{{Ford} \& {Rasio}(2008)}]{Ford_and_Rasio_2008}
{Ford}, E.~B., \& {Rasio}, F.~A. 2008, \apj, 686, 621

\bibitem[{{Ford} {et~al.}(2003){Ford}, {Rasio}, \& {Yu}}]{Ford_et_al_2003}
{Ford}, E.~B., {Rasio}, F.~A., \& {Yu}, K. 2003, in Astronomical Society of the
  Pacific Conference Series, Vol. 294, Scientific Frontiers in Research on
  Extrasolar Planets, ed. D.~{Deming} \& S.~{Seager}, 181--188

\bibitem[{{Fortney} \& {Hubbard}(2004)}]{Fortney_and_Hubbard_2004}
{Fortney}, J.~J., \& {Hubbard}, W.~B. 2004, \apj, 608, 1039

\bibitem[{{Fortney} {et~al.}(2008){Fortney}, {Lodders}, {Marley}, \&
  {Freedman}}]{Fortney_et_al_2008_1}
{Fortney}, J.~J., {Lodders}, K., {Marley}, M.~S., \& {Freedman}, R.~S. 2008,
  \apj, 678, 1419

\bibitem[{{Fortney} {et~al.}(2007){Fortney}, {Marley}, \&
  {Barnes}}]{Fortney_et_al_2007}
{Fortney}, J.~J., {Marley}, M.~S., \& {Barnes}, J.~W. 2007, \apj, 659, 1661

\bibitem[{{Gillon} {et~al.}(2009{\natexlab{a}}){Gillon}, {Anderson}, {Triaud},
  {Hellier}, {Maxted}, {Pollaco}, {Queloz}, {Smalley}, {West}, {Wilson},
  {Bentley}, {Collier Cameron}, {Enoch}, {Hebb}, {Horne}, {Irwin}, {Joshi},
  {Lister}, {Mayor}, {Pepe}, {Parley}, {Segransan}, {Udry}, \&
  {Wheatley}}]{Gillon_et_al_2009_2}
{Gillon}, M., {Anderson}, D.~R., {Triaud}, A.~H.~M.~J., {Hellier}, C.,
  {Maxted}, P.~F.~L., {Pollaco}, D., {Queloz}, D., {Smalley}, B., {West},
  R.~G., {Wilson}, D.~M., {Bentley}, S.~J., {Collier Cameron}, A., {Enoch}, B.,
  {Hebb}, L., {Horne}, K., {Irwin}, J., {Joshi}, Y.~C., {Lister}, T.~A.,
  {Mayor}, M., {Pepe}, F., {Parley}, N., {Segransan}, D., {Udry}, S., \&
  {Wheatley}, P.~J. 2009{\natexlab{a}}, \aap, 501, 785

\bibitem[{{Gillon} {et~al.}(2009{\natexlab{b}}){Gillon}, {Demory}, {Triaud},
  {Barman}, {Hebb}, {Montalb{\'a}n}, {Maxted}, {Queloz}, {Deleuil}, \&
  {Magain}}]{Gillon_et_al_2009_3}
{Gillon}, M., {Demory}, B., {Triaud}, A.~H.~M.~J., {Barman}, T., {Hebb}, L.,
  {Montalb{\'a}n}, J., {Maxted}, P.~F.~L., {Queloz}, D., {Deleuil}, M., \&
  {Magain}, P. 2009{\natexlab{b}}, \aap, 506, 359

\bibitem[{{Gillon} {et~al.}(2009{\natexlab{c}}){Gillon}, {Smalley}, {Hebb},
  {Anderson}, {Triaud}, {Hellier}, {Maxted}, {Queloz}, \&
  {Wilson}}]{Gillon_et_al_2009_1}
{Gillon}, M., {Smalley}, B., {Hebb}, L., {Anderson}, D.~R., {Triaud},
  A.~H.~M.~J., {Hellier}, C., {Maxted}, P.~F.~L., {Queloz}, D., \& {Wilson},
  D.~M. 2009{\natexlab{c}}, \aap, 496, 259

\bibitem[{{Goldreich} \& {Sari}(2003)}]{Goldreich_and_Sari_2003}
{Goldreich}, P., \& {Sari}, R. 2003, \apj, 585, 1024

\bibitem[{{Goldreich} \& {Soter}(1966)}]{Goldreich+Soter_1966}
{Goldreich}, P., \& {Soter}, S. 1966, Icarus, 5, 375

\bibitem[{{Goldreich}(1963)}]{Goldreich_1963}
{Goldreich}, R. 1963, \mnras, 126, 257

\bibitem[{{Goodman} \& {Lackner}(2009)}]{Goodman_and_Lackner_2009}
{Goodman}, J., \& {Lackner}, C. 2009, \apj, 696, 2054

\bibitem[{{Gu} {et~al.}(2004){Gu}, {Bodenheimer}, \& {Lin}}]{Gu_et_al_2004}
{Gu}, P.-G., {Bodenheimer}, P.~H., \& {Lin}, D.~N.~C. 2004, \apj, 608, 1076

\bibitem[{{Gu} {et~al.}(2003){Gu}, {Lin}, \& {Bodenheimer}}]{Gu_et_al_2003}
{Gu}, P.-G., {Lin}, D.~N.~C., \& {Bodenheimer}, P.~H. 2003, \apj, 588, 509

\bibitem[{{Guillot}(2008)}]{Guillot_2008}
{Guillot}, T. 2008, Physica Scripta Volume T, 130, 014023

\bibitem[{{Guillot} {et~al.}(1996){Guillot}, {Burrows}, {Hubbard}, {Lunine}, \&
  {Saumon}}]{guillot_et_al1996}
{Guillot}, T., {Burrows}, A., {Hubbard}, W.~B., {Lunine}, J.~I., \& {Saumon},
  D. 1996, \apjl, 459, L35

\bibitem[{{Guillot} \& {Gladman}(2000)}]{Guillot_and_Gladman_2000}
{Guillot}, T., \& {Gladman}, B. 2000, in Astronomical Society of the Pacific
  Conference Series, Vol. 219, Disks, Planetesimals, and Planets, ed.
  G.~{Garz{\'o}n}, C.~{Eiroa}, D.~{de Winter}, \& T.~J. {Mahoney}, 475

\bibitem[{{Guillot} {et~al.}(2006){Guillot}, {Santos}, {Pont}, {Iro}, {Melo},
  \& {Ribas}}]{guillot06}
{Guillot}, T., {Santos}, N.~C., {Pont}, F., {Iro}, N., {Melo}, C., \& {Ribas},
  I. 2006, \aap, 453, L21

\bibitem[{{Hartman}(2009)}]{Hartman_2009}
{Hartman}, J.~D. 2009, private communication

\bibitem[{{Hartman} {et~al.}(2009){Hartman}, {Bakos}, {Torres}, {Kov{\'a}cs},
  {Noyes}, {P{\'a}l}, {Latham}, {Sip{\H o}cz}, {Fischer}, {Johnson}, {Marcy},
  {Butler}, {Howard}, {Esquerdo}, {Sasselov}, {Kov{\'a}cs}, {Stefanik},
  {Fernandez}, {L{\'a}z{\'a}r}, {Papp}, \& {S{\'a}ri}}]{Hartman_et_al_2009}
{Hartman}, J.~D., {Bakos}, G.~{\'A}., {Torres}, G., {Kov{\'a}cs}, G., {Noyes},
  R.~W., {P{\'a}l}, A., {Latham}, D.~W., {Sip{\H o}cz}, B., {Fischer}, D.~A.,
  {Johnson}, J.~A., {Marcy}, G.~W., {Butler}, R.~P., {Howard}, A.~W.,
  {Esquerdo}, G.~A., {Sasselov}, D.~D., {Kov{\'a}cs}, G., {Stefanik}, R.~P.,
  {Fernandez}, J.~M., {L{\'a}z{\'a}r}, J., {Papp}, I., \& {S{\'a}ri}, P. 2009,
  \apj, 706, 785

\bibitem[{{Hebb} {et~al.}(2009){Hebb}, {Collier-Cameron}, {Loeillet},
  {Pollacco}, {H{\'e}brard}, {Street}, {Bouchy}, {Stempels}, {Moutou},
  {Simpson}, {Udry}, {Joshi}, {West}, {Skillen}, {Wilson}, {McDonald},
  {Gibson}, {Aigrain}, {Anderson}, {Benn}, {Christian}, {Enoch}, {Haswell},
  {Hellier}, {Horne}, {Irwin}, {Lister}, {Maxted}, {Mayor}, {Norton}, {Parley},
  {Pont}, {Queloz}, {Smalley}, \& {Wheatley}}]{Hebb_et_al_2009}
{Hebb}, L., {Collier-Cameron}, A., {Loeillet}, B., {Pollacco}, D.,
  {H{\'e}brard}, G., {Street}, R.~A., {Bouchy}, F., {Stempels}, H.~C.,
  {Moutou}, C., {Simpson}, E., {Udry}, S., {Joshi}, Y.~C., {West}, R.~G.,
  {Skillen}, I., {Wilson}, D.~M., {McDonald}, I., {Gibson}, N.~P., {Aigrain},
  S., {Anderson}, D.~R., {Benn}, C.~R., {Christian}, D.~J., {Enoch}, B.,
  {Haswell}, C.~A., {Hellier}, C., {Horne}, K., {Irwin}, J., {Lister}, T.~A.,
  {Maxted}, P., {Mayor}, M., {Norton}, A.~J., {Parley}, N., {Pont}, F.,
  {Queloz}, D., {Smalley}, B., \& {Wheatley}, P.~J. 2009, \apj, 693, 1920

\bibitem[{{Henry} {et~al.}(2000){Henry}, {Marcy}, {Butler}, \&
  {Vogt}}]{henry_et_al2000}
{Henry}, G.~W., {Marcy}, G.~W., {Butler}, R.~P., \& {Vogt}, S.~S. 2000, \apjl,
  529, L41

\bibitem[{{Hubeny}(1988)}]{Hubeny_1988}
{Hubeny}, I. 1988, Computer Physics Communications, 52, 103

\bibitem[{{Hubeny} {et~al.}(2003){Hubeny}, {Burrows}, \&
  {Sudarsky}}]{hubeny_et_al2003}
{Hubeny}, I., {Burrows}, A., \& {Sudarsky}, D. 2003, \apj, 594, 1011

\bibitem[{{Hubeny} \& {Lanz}(1995)}]{Hubeny_and_Lanz_1995}
{Hubeny}, I., \& {Lanz}, T. 1995, \apj, 439, 875

\bibitem[{{Ibgui} \& {Burrows}(2009)}]{Ibgui_and_Burrows_2009}
{Ibgui}, L., \& {Burrows}, A. 2009, \apj, 700, 1921

\bibitem[{{Ibgui} {et~al.}(2009){Ibgui}, {Spiegel}, \&
  {Burrows}}]{Ibgui_et_al_2009_2}
{Ibgui}, L., {Spiegel}, D.~S., \& {Burrows}, A. 2009, submitted to \apj,
  arXiv:0910.5928

\bibitem[{{Ikoma} {et~al.}(2006){Ikoma}, {Guillot}, {Genda}, {Tanigawa}, \&
  {Ida}}]{Ikoma_et_al_2006}
{Ikoma}, M., {Guillot}, T., {Genda}, H., {Tanigawa}, T., \& {Ida}, S. 2006,
  \apj, 650, 1150

\bibitem[{{Jackson} {et~al.}(2008){Jackson}, {Greenberg}, \&
  {Barnes}}]{Jackson_et_al_2008_3}
{Jackson}, B., {Greenberg}, R., \& {Barnes}, R. 2008, \apj, 681, 1631

\bibitem[{{Johns-Krull} {et~al.}(2008){Johns-Krull}, {McCullough}, {Burke},
  {Valenti}, {Janes}, {Heasley}, {Prato}, {Bissinger}, {Fleenor}, {Foote},
  {Garcia-Melendo}, {Gary}, {Howell}, {Mallia}, {Masi}, \&
  {Vanmunster}}]{Johns-Krull_et_al_2008}
{Johns-Krull}, C.~M., {McCullough}, P.~R., {Burke}, C.~J., {Valenti}, J.~A.,
  {Janes}, K.~A., {Heasley}, J.~N., {Prato}, L., {Bissinger}, R., {Fleenor},
  M., {Foote}, C.~N., {Garcia-Melendo}, E., {Gary}, B.~L., {Howell}, P.~J.,
  {Mallia}, F., {Masi}, G., \& {Vanmunster}, T. 2008, \apj, 677, 657

\bibitem[{{Johnson} {et~al.}(2008){Johnson}, {Winn}, {Narita}, {Enya},
  {Williams}, {Marcy}, {Sato}, {Ohta}, {Taruya}, {Suto}, {Turner}, {Bakos},
  {Butler}, {Vogt}, {Aoki}, {Tamura}, {Yamada}, {Yoshii}, \&
  {Hidas}}]{Johnson_et_al_2008}
{Johnson}, J.~A., {Winn}, J.~N., {Narita}, N., {Enya}, K., {Williams},
  P.~K.~G., {Marcy}, G.~W., {Sato}, B., {Ohta}, Y., {Taruya}, A., {Suto}, Y.,
  {Turner}, E.~L., {Bakos}, G., {Butler}, R.~P., {Vogt}, S.~S., {Aoki}, W.,
  {Tamura}, M., {Yamada}, T., {Yoshii}, Y., \& {Hidas}, M. 2008, \apj, 686, 649

\bibitem[{{Juri{\'c}} \& {Tremaine}(2008)}]{Juric_and_Tremaine_2008}
{Juri{\'c}}, M., \& {Tremaine}, S. 2008, \apj, 686, 603

\bibitem[{{Knutson} {et~al.}(2008){Knutson}, {Charbonneau}, {Allen}, {Burrows},
  \& {Megeath}}]{Knutson_et_al_2008}
{Knutson}, H.~A., {Charbonneau}, D., {Allen}, L.~E., {Burrows}, A., \&
  {Megeath}, S.~T. 2008, \apj, 673, 526

\bibitem[{{Knutson} {et~al.}(2009){Knutson}, {Charbonneau}, {Burrows},
  {O'Donovan}, \& {Mandushev}}]{Knutson_et_al_2009_1}
{Knutson}, H.~A., {Charbonneau}, D., {Burrows}, A., {O'Donovan}, F.~T., \&
  {Mandushev}, G. 2009, \apj, 691, 866

\bibitem[{{Knutson} {et~al.}(2007){Knutson}, {Charbonneau}, {Noyes}, {Brown},
  \& {Gilliland}}]{knutson_et_al2007a}
{Knutson}, H.~A., {Charbonneau}, D., {Noyes}, R.~W., {Brown}, T.~M., \&
  {Gilliland}, R.~L. 2007, \apj, 655, 564

\bibitem[{{Kurucz}(1994)}]{Kurucz_1994}
{Kurucz}, R. 1994, Solar abundance model atmospheres for 0,1,2,4,8 km/s.~Kurucz
  CD-ROM No.~19.~ Cambridge, Mass.: Smithsonian Astrophysical Observatory,
  1994., 19

\bibitem[{{Laughlin} {et~al.}(2005){Laughlin}, {Wolf}, {Vanmunster},
  {Bodenheimer}, {Fischer}, {Marcy}, {Butler}, \&
  {Vogt}}]{laughlin_et_al_2005_1}
{Laughlin}, G., {Wolf}, A., {Vanmunster}, T., {Bodenheimer}, P., {Fischer}, D.,
  {Marcy}, G., {Butler}, P., \& {Vogt}, S. 2005, \apj, 621, 1072

\bibitem[{{Leconte} {et~al.}(2009){Leconte}, {Baraffe}, {Chabrier}, {Barman},
  \& {Levrard}}]{Leconte_et_al_2009}
{Leconte}, J., {Baraffe}, I., {Chabrier}, G., {Barman}, T., \& {Levrard}, B.
  2009, \aap, 506, 385

\bibitem[{{Li} {et~al.}(2009){Li}, {Miller}, {Lin}, \&
  {Fortney}}]{Li_et_al_2009}
{Li}, S., {Miller}, N., {Lin}, D., \& {Fortney}, J. 2009, \nat, submitted

\bibitem[{{Lin} {et~al.}(1996){Lin}, {Bodenheimer}, \&
  {Richardson}}]{Lin_et_al_1996}
{Lin}, D.~N.~C., {Bodenheimer}, P., \& {Richardson}, D.~C. 1996, \nat, 380, 606

\bibitem[{{Liu} {et~al.}(2008){Liu}, {Burrows}, \& {Ibgui}}]{Liu_et_al_2008}
{Liu}, X., {Burrows}, A., \& {Ibgui}, L. 2008, \apj, 687, 1191

\bibitem[{{Madhusudhan} \& {Winn}(2009)}]{Madhusudhan_and_Winn_2009}
{Madhusudhan}, N., \& {Winn}, J.~N. 2009, \apj, 693, 784

\bibitem[{{Mandushev} {et~al.}(2007){Mandushev}, {O'Donovan}, {Charbonneau},
  {Torres}, {Latham}, {Bakos}, {Dunham}, {Sozzetti}, {Fern{\'a}ndez},
  {Esquerdo}, {Everett}, {Brown}, {Rabus}, {Belmonte}, \&
  {Hillenbrand}}]{Mandushev_et_al_2007}
{Mandushev}, G., {O'Donovan}, F.~T., {Charbonneau}, D., {Torres}, G., {Latham},
  D.~W., {Bakos}, G.~{\'A}., {Dunham}, E.~W., {Sozzetti}, A., {Fern{\'a}ndez},
  J.~M., {Esquerdo}, G.~A., {Everett}, M.~E., {Brown}, T.~M., {Rabus}, M.,
  {Belmonte}, J.~A., \& {Hillenbrand}, L.~A. 2007, \apjl, 667, L195

\bibitem[{{Mardling}(2007)}]{Mardling_2007}
{Mardling}, R.~A. 2007, \mnras, 382, 1768

\bibitem[{{Marley} {et~al.}(2007){Marley}, {Fortney}, {Hubickyj},
  {Bodenheimer}, \& {Lissauer}}]{Marley_et_al_2007}
{Marley}, M.~S., {Fortney}, J.~J., {Hubickyj}, O., {Bodenheimer}, P., \&
  {Lissauer}, J.~J. 2007, \apj, 655, 541

\bibitem[{{Miller} {et~al.}(2009){Miller}, {Fortney}, \&
  {Jackson}}]{Miller_et_al_2009}
{Miller}, N., {Fortney}, J.~J., \& {Jackson}, B. 2009, \apj, 702, 1413

\bibitem[{{Murray} \& {Dermott}(1999)}]{Murray_et_Dermott_1999}
{Murray}, C.~D., \& {Dermott}, S.~F. 1999, {Solar System Dynamics}, ed.
  C.~U.~P. (MD99)

\bibitem[{{Nagasawa} {et~al.}(2008){Nagasawa}, {Ida}, \&
  {Bessho}}]{Nagasawa_et_al_2008}
{Nagasawa}, M., {Ida}, S., \& {Bessho}, T. 2008, \apj, 678, 498

\bibitem[{{O'Donovan} {et~al.}(2006){O'Donovan}, {Charbonneau}, {Mandushev},
  {Dunham}, {Latham}, {Torres}, {Sozzetti}, {Brown}, {Trauger}, {Belmonte},
  {Rabus}, {Almenara}, {Alonso}, {Deeg}, {Esquerdo}, {Falco}, {Hillenbrand},
  {Roussanova}, {Stefanik}, \& {Winn}}]{odonovan_et_al2006b}
{O'Donovan}, F.~T., {Charbonneau}, D., {Mandushev}, G., {Dunham}, E.~W.,
  {Latham}, D.~W., {Torres}, G., {Sozzetti}, A., {Brown}, T.~M., {Trauger},
  J.~T., {Belmonte}, J.~A., {Rabus}, M., {Almenara}, J.~M., {Alonso}, R.,
  {Deeg}, H.~J., {Esquerdo}, G.~A., {Falco}, E.~E., {Hillenbrand}, L.~A.,
  {Roussanova}, A., {Stefanik}, R.~P., \& {Winn}, J.~N. 2006, \apjl, 651, L61

\bibitem[{{Ogilvie} \& {Lin}(2004)}]{Ogilvie_and_Lin_2004}
{Ogilvie}, G.~I., \& {Lin}, D.~N.~C. 2004, \apj, 610, 477

\bibitem[{{P{\'a}l} {et~al.}(2008){P{\'a}l}, {Bakos}, {Torres}, {Noyes},
  {Latham}, {Kov{\'a}cs}, {Marcy}, {Fischer}, {Butler}, {Sasselov}, {Sip{\H
  o}cz}, {Esquerdo}, {Kov{\'a}cs}, {Stefanik}, {L{\'a}z{\'a}r}, {Papp}, \&
  {S{\'a}ri}}]{Pal_et_al_2008}
{P{\'a}l}, A., {Bakos}, G.~{\'A}., {Torres}, G., {Noyes}, R.~W., {Latham},
  D.~W., {Kov{\'a}cs}, G., {Marcy}, G.~W., {Fischer}, D.~A., {Butler}, R.~P.,
  {Sasselov}, D.~D., {Sip{\H o}cz}, B., {Esquerdo}, G.~A., {Kov{\'a}cs}, G.,
  {Stefanik}, R., {L{\'a}z{\'a}r}, J., {Papp}, I., \& {S{\'a}ri}, P. 2008,
  \apj, 680, 1450

\bibitem[{{Peale} \& {Cassen}(1978)}]{Peale_and_Cassen_1978}
{Peale}, S.~J., \& {Cassen}, P. 1978, Icarus, 36, 245

\bibitem[{{Pont} {et~al.}(2009){Pont}, {H{\'e}brard}, {Irwin}, {Bouchy},
  {Moutou}, {Ehrenreich}, {Guillot}, {Aigrain}, {Bonfils}, {Berta}, {Boisse},
  {Burke}, {Charbonneau}, {Delfosse}, {Desort}, {Eggenberger}, {Forveille},
  {Lagrange}, {Lovis}, {Nutzman}, {Pepe}, {Perrier}, {Queloz}, {Santos},
  {S{\'e}gransan}, {Udry}, \& {Vidal-Madjar}}]{Pont_et_al_2009_1}
{Pont}, F., {H{\'e}brard}, G., {Irwin}, J.~M., {Bouchy}, F., {Moutou}, C.,
  {Ehrenreich}, D., {Guillot}, T., {Aigrain}, S., {Bonfils}, X., {Berta}, Z.,
  {Boisse}, I., {Burke}, C., {Charbonneau}, D., {Delfosse}, X., {Desort}, M.,
  {Eggenberger}, A., {Forveille}, T., {Lagrange}, A., {Lovis}, C., {Nutzman},
  P., {Pepe}, F., {Perrier}, C., {Queloz}, D., {Santos}, N.~C.,
  {S{\'e}gransan}, D., {Udry}, S., \& {Vidal-Madjar}, A. 2009, \aap, 502, 695

\bibitem[{{Poynting}(1903)}]{poynting1903}
{Poynting}, J.~H. 1903, \mnras, 64, A1

\bibitem[{{Rauer} {et~al.}(2009){Rauer}, {Queloz}, {Csizmadia}, {Deleuil},
  {Alonso}, {Aigrain}, {Almenara}, {Auvergne}, {Baglin}, {Barge}, {Bord{\'e}},
  {Bouchy}, {Bruntt}, {Cabrera}, {Carone}, {Carpano}, {de La Reza}, {Deeg},
  {Dvorak}, {Erikson}, {Fridlund}, {Gandolfi}, {Gillon}, {Guillot}, {Guenther},
  {Hatzes}, {H{\'e}brard}, {Kabath}, {Jorda}, {Lammer}, {L{\'e}ger},
  {Llebaria}, {Magain}, {Mazeh}, {Moutou}, {Ollivier}, {P{\"a}tzold}, {Pont},
  {Rabus}, {Renner}, {Rouan}, {Shporer}, {Samuel}, {Schneider}, {Triaud}, \&
  {Wuchterl}}]{Rauer_et_al_2009}
{Rauer}, H., {Queloz}, D., {Csizmadia}, S., {Deleuil}, M., {Alonso}, R.,
  {Aigrain}, S., {Almenara}, J.~M., {Auvergne}, M., {Baglin}, A., {Barge}, P.,
  {Bord{\'e}}, P., {Bouchy}, F., {Bruntt}, H., {Cabrera}, J., {Carone}, L.,
  {Carpano}, S., {de La Reza}, R., {Deeg}, H.~J., {Dvorak}, R., {Erikson}, A.,
  {Fridlund}, M., {Gandolfi}, D., {Gillon}, M., {Guillot}, T., {Guenther}, E.,
  {Hatzes}, A., {H{\'e}brard}, G., {Kabath}, P., {Jorda}, L., {Lammer}, H.,
  {L{\'e}ger}, A., {Llebaria}, A., {Magain}, P., {Mazeh}, T., {Moutou}, C.,
  {Ollivier}, M., {P{\"a}tzold}, M., {Pont}, F., {Rabus}, M., {Renner}, S.,
  {Rouan}, D., {Shporer}, A., {Samuel}, B., {Schneider}, J., {Triaud},
  A.~H.~M.~J., \& {Wuchterl}, G. 2009, \aap, 506, 281

\bibitem[{{Robertson}(1937)}]{robertson1937}
{Robertson}, H.~P. 1937, \mnras, 97, 423

\bibitem[{{Saumon} {et~al.}(1995){Saumon}, {Chabrier}, \& {van
  Horn}}]{Saumon_et_al_1995}
{Saumon}, D., {Chabrier}, G., \& {van Horn}, H.~M. 1995, \apjs, 99, 713

\bibitem[{{Shen} \& {Turner}(2008)}]{Shen_and_Turner_2008}
{Shen}, Y., \& {Turner}, E.~L. 2008, \apj, 685, 553

\bibitem[{{Southworth} {et~al.}(2009){Southworth}, {Hinse}, {Burgdorf},
  {Dominik}, {Hornstrup}, {J{\o}rgensen}, {Liebig}, {Ricci}, {Th{\"o}ne},
  {Anguita}, {Bozza}, {Novati}, {Harps{\o}e}, {Mancini}, {Masi}, {Mathiasen},
  {Rahvar}, {Scarpetta}, {Snodgrass}, {Surdej}, \&
  {Zub}}]{Southworth_et_al_2009_2}
{Southworth}, J., {Hinse}, T.~C., {Burgdorf}, M.~J., {Dominik}, M.,
  {Hornstrup}, A., {J{\o}rgensen}, U.~G., {Liebig}, C., {Ricci}, D.,
  {Th{\"o}ne}, C.~C., {Anguita}, T., {Bozza}, V., {Novati}, S.~C.,
  {Harps{\o}e}, K., {Mancini}, L., {Masi}, G., {Mathiasen}, M., {Rahvar}, S.,
  {Scarpetta}, G., {Snodgrass}, C., {Surdej}, J., \& {Zub}, M. 2009, \mnras,
  399, 287

\bibitem[{{Sozzetti} {et~al.}(2009){Sozzetti}, {Torres}, {Charbonneau}, {Winn},
  {Korzennik}, {Holman}, {Latham}, {Laird}, {Fernandez}, {O'Donovan},
  {Mandushev}, {Dunham}, {Everett}, {Esquerdo}, {Rabus}, {Belmonte}, {Deeg},
  {Brown}, {Hidas}, \& {Baliber}}]{Sozzetti_et_al_2008}
{Sozzetti}, A., {Torres}, G., {Charbonneau}, D., {Winn}, J.~N., {Korzennik},
  S.~G., {Holman}, M.~J., {Latham}, D.~W., {Laird}, J.~B., {Fernandez}, J.,
  {O'Donovan}, F.~T., {Mandushev}, G., {Dunham}, E., {Everett}, M.~E.,
  {Esquerdo}, G.~A., {Rabus}, M., {Belmonte}, J.~A., {Deeg}, H.~J., {Brown},
  T.~N., {Hidas}, M.~G., \& {Baliber}, N. 2009, \apj, 691, 1145

\bibitem[{{Spiegel} {et~al.}(2010){Spiegel}, {Goodman}, {Burrows}, \&
  {Ibgui}}]{Spiegel_et_al_2010_2}
{Spiegel}, D.~S., {Goodman}, J., {Burrows}, A., \& {Ibgui}, L. 2010, \apj, in
  preparation

\bibitem[{{Spiegel} {et~al.}(2009){Spiegel}, {Silverio}, \&
  {Burrows}}]{Spiegel_et_al_2009_1}
{Spiegel}, D.~S., {Silverio}, K., \& {Burrows}, A. 2009, \apj, 699, 1487

\bibitem[{{Thompson} \& {Lauson}(1972)}]{Thompson_and_Lauson_1972}
{Thompson}, S.~L., \& {Lauson}, H.~S. 1972, Sandia National Laboratory Report,
  SC-RR-71,0714

\bibitem[{{West} {et~al.}(2009){West}, {Anderson}, {Gillon}, {Hebb}, {Hellier},
  {Maxted}, {Queloz}, {Smalley}, {Triaud}, {Wilson}, {Bentley}, {Collier
  Cameron}, {Enoch}, {Horne}, {Irwin}, {Lister}, {Mayor}, {Parley}, {Pepe},
  {Pollacco}, {Segransan}, {Spano}, {Udry}, \& {Wheatley}}]{West_et_al_2009}
{West}, R.~G., {Anderson}, D.~R., {Gillon}, M., {Hebb}, L., {Hellier}, C.,
  {Maxted}, P.~F.~L., {Queloz}, D., {Smalley}, B., {Triaud}, A.~H.~M.~J.,
  {Wilson}, D.~M., {Bentley}, S.~J., {Collier Cameron}, A., {Enoch}, B.,
  {Horne}, K., {Irwin}, J., {Lister}, T.~A., {Mayor}, M., {Parley}, N., {Pepe},
  F., {Pollacco}, D., {Segransan}, D., {Spano}, M., {Udry}, S., \& {Wheatley},
  P.~J. 2009, \aj, 137, 4834

\bibitem[{{Wilson} {et~al.}(2008){Wilson}, {Gillon}, {Hellier}, {Maxted},
  {Pepe}, {Queloz}, {Anderson}, {Collier Cameron}, {Smalley}, {Lister},
  {Bentley}, {Blecha}, {Christian}, {Enoch}, {Haswell}, {Hebb}, {Horne},
  {Irwin}, {Joshi}, {Kane}, {Marmier}, {Mayor}, {Parley}, {Pollacco}, {Pont},
  {Ryans}, {Segransan}, {Skillen}, {Street}, {Udry}, {West}, \&
  {Wheatley}}]{Wilson_et_al_2008}
{Wilson}, D.~M., {Gillon}, M., {Hellier}, C., {Maxted}, P.~F.~L., {Pepe}, F.,
  {Queloz}, D., {Anderson}, D.~R., {Collier Cameron}, A., {Smalley}, B.,
  {Lister}, T.~A., {Bentley}, S.~J., {Blecha}, A., {Christian}, D.~J., {Enoch},
  B., {Haswell}, C.~A., {Hebb}, L., {Horne}, K., {Irwin}, J., {Joshi}, Y.~C.,
  {Kane}, S.~R., {Marmier}, M., {Mayor}, M., {Parley}, N., {Pollacco}, D.,
  {Pont}, F., {Ryans}, R., {Segransan}, D., {Skillen}, I., {Street}, R.~A.,
  {Udry}, S., {West}, R.~G., \& {Wheatley}, P.~J. 2008, \apjl, 675, L113

\bibitem[{{Winn} {et~al.}(2007){Winn}, {Holman}, {Bakos}, {P{\'a}l}, {Johnson},
  {Williams}, {Shporer}, {Mazeh}, {Fernandez}, {Latham}, \&
  {Gillon}}]{Winn_et_al_2007}
{Winn}, J.~N., {Holman}, M.~J., {Bakos}, G.~{\'A}., {P{\'a}l}, A., {Johnson},
  J.~A., {Williams}, P.~K.~G., {Shporer}, A., {Mazeh}, T., {Fernandez}, J.,
  {Latham}, D.~W., \& {Gillon}, M. 2007, \aj, 134, 1707

\bibitem[{{Winn} {et~al.}(2009{\natexlab{a}}){Winn}, {Holman}, {Carter},
  {Torres}, {Osip}, \& {Beatty}}]{Winn_et_al_2009_1}
{Winn}, J.~N., {Holman}, M.~J., {Carter}, J.~A., {Torres}, G., {Osip}, D.~J.,
  \& {Beatty}, T. 2009{\natexlab{a}}, \aj, 137, 3826

\bibitem[{{Winn} {et~al.}(2008){Winn}, {Holman}, {Torres}, {McCullough},
  {Johns-Krull}, {Latham}, {Shporer}, {Mazeh}, {Garcia-Melendo}, {Foote},
  {Esquerdo}, \& {Everett}}]{Winn_et_al_2008}
{Winn}, J.~N., {Holman}, M.~J., {Torres}, G., {McCullough}, P., {Johns-Krull},
  C., {Latham}, D.~W., {Shporer}, A., {Mazeh}, T., {Garcia-Melendo}, E.,
  {Foote}, C., {Esquerdo}, G., \& {Everett}, M. 2008, \apj, 683, 1076

\bibitem[{{Winn} {et~al.}(2009{\natexlab{b}}){Winn}, {Johnson}, {Albrecht},
  {Howard}, {Marcy}, {Crossfield}, \& {Holman}}]{Winn_et_al_2009_4}
{Winn}, J.~N., {Johnson}, J.~A., {Albrecht}, S., {Howard}, A.~W., {Marcy},
  G.~W., {Crossfield}, I.~J., \& {Holman}, M.~J. 2009{\natexlab{b}}, \apjl,
  703, L99

\bibitem[{{Wright}(2009)}]{Wright_2009}
{Wright}, J.~T. 2009, arXiv:0909.0957

\bibitem[{{Wu}(2003)}]{Wu_2003}
{Wu}, Y. 2003, in Astronomical Society of the Pacific Conference Series, Vol.
  294, Scientific Frontiers in Research on Extrasolar Planets, ed. D.~{Deming}
  \& S.~{Seager}, 213--216

\bibitem[{{Wu}(2005)}]{Wu_2005_1}
{Wu}, Y. 2005, \apj, 635, 674

\bibitem[{{Wu} \& {Murray}(2003)}]{Wu_and_Murray_2003}
{Wu}, Y., \& {Murray}, N. 2003, \apj, 589, 605

\bibitem[{{Wu} {et~al.}(2007){Wu}, {Murray}, \& {Ramsahai}}]{Wu_et_al_2007}
{Wu}, Y., {Murray}, N.~W., \& {Ramsahai}, J.~M. 2007, \apj, 670, 820

\bibitem[{{Yoder} \& {Peale}(1981)}]{Yoder_and_Peale_1981}
{Yoder}, C.~F., \& {Peale}, S.~J. 1981, Icarus, 47, 1

\end{thebibliography}

\clearpage
\begin{table}[ht]
\small
\begin{center}
\caption{Planet and Orbital Data} \label{tab:transit_planets_data}
\vspace{0.4in}
\begin{tabular}{lccccccccc}
\hline\hline
\rule {0pt} {10pt}
  Planet          &            $a$                   &         $e$                  &    $P$        &         \multicolumn{2} {c}{$M_{p}$}                     &         $R_p$            &  $F_{p}$$^a$      & Roche limit &     References \\
                  &            (AU)                  &                              &   (days)      &        ($\rm M_{J}$)        &    ($\rm M_{\earth}$)      &      ($\rm R_{J}$)         &                   &    (AU)     &                \\
  \tableline \\[0.0cm]
   HAT-P-12b      &  $0.0384 ^{+0.0003 }_{-0.0003 }$ & $< 0.065               $$^b$ &   $3.21306$   &  $0.211^{+0.012}_{-0.012}$  &  $67.1^{+3.8 }_{-3.8 }$    &  $0.959^{+0.029}_{-0.021}$ &    $0.191$        & 0.017       &       1,2      \\[0.2cm]
   TrES-4         &  $0.05105^{+0.00079}_{-0.00167}$ & $< 0.01                $$^c$ &   $3.55395$   &  $0.925^{+0.081}_{-0.082}$  &  $294 ^{+25.7}_{-26.1}$    &  $1.783^{+0.093}_{-0.086}$ &    $2.371$        & 0.024       &        3       \\[0.2cm]
   WASP-4b        &  $0.02340^{+0.00060}_{-0.00060}$ & $< 0.096                   $ &   $1.33823$   &  $1.237^{+0.064}_{-0.064}$  &  $393 ^{+20.3}_{-20.3}$    &  $1.365^{+0.021}_{-0.021}$ &    $1.706$        & 0.015       &       4,5      \\[0.2cm]
   WASP-6b        &  $0.0421 ^{+0.0008 }_{-0.0013 }$ & $  0.054^{+0.018}_{-0.015} $ &   $3.36101$   &  $0.503^{+0.019}_{-0.038}$  &  $160 ^{+6.0 }_{-12.1}$    &  $1.224^{+0.051}_{-0.052}$ &    $0.463$        & 0.018       &        6       \\[0.2cm]
   WASP-12b       &  $0.0229 ^{+0.0008 }_{-0.0008 }$ & $  0.049^{+0.015}_{-0.015} $ &   $1.09142$   &  $1.41 ^{+0.10 }_{-0.10 }$  &  $448 ^{+31.8}_{-31.8}$    &  $1.79 ^{+0.09} _{-0.09} $ &    $9.098$        & 0.021       &        7       \\[0.2cm]
   WASP-15b       &  $0.0499 ^{+0.0018 }_{-0.0018 }$ & $  0.052^{+0.029}_{-0.040} $ &   $3.75207$   &  $0.542^{+0.050}_{-0.050}$  &  $172 ^{+15.9}_{-15.9}$    &  $1.428^{+0.077}_{-0.077}$ &    $1.696$        & 0.022       &        8       \\[0.2cm]
  \tableline
\end{tabular}
\bigskip
\tablenotetext{}{$^{a}$ $F_{p}$ is the stellar flux at the planet's substellar point, in units of $\rm 10^{9}~ergs~cm^{-2}~s^{-1}$.}
\tablenotetext{}{$^{b}$ According to \citet{Hartman_2009}, the 68.3$\%$ confidence upper limit is 0.065 and the 95.4$\%$ confidence upper limit is 0.085.}
\tablenotetext{}{$^{c}$ For specificity, we adopt this value of $0.01$. 
                 \citet{Knutson_et_al_2009_1} derive a $3\sigma$ upper limit $|e\cos(\omega)|$ of ~0.0058, where $\omega$ is the argument of periastron.}
\tablerefs{ (1) \citet{Hartman_et_al_2009},        
            (2) \citet{Hartman_2009},              
            (3) \citet{Sozzetti_et_al_2008},       
            (4) \citet{Winn_et_al_2009_1},         
            (5) \citet{Madhusudhan_and_Winn_2009}, 
            (6) \citet{Gillon_et_al_2009_2},       
            (7) \citet{Hebb_et_al_2009},           
	    (8) \citet{West_et_al_2009}.           
}
\end{center}
\end{table}

\vspace{2cm}

\begin{table}[ht]
\small
\begin{center}
\caption{Host Star Data} \label{tab:host_stars_data}
\vspace{0.2in}

\begin{tabular}{p{3cm}ccccc}
\hline\hline
\rule {0pt} {10pt}
  Star               &          $M_{\ast}$           &            $R_{\ast}$           &        $T_{\ast}$         &        $\left[ Fe/H \right]_{\ast}$     &        Age          \\
                     &        $\rm M_{\sun}$         &          $\rm R_{\sun}$         &          (K)              &                (dex)                    &       (Gyr)         \\
  \tableline \\[0.0cm]
   HAT-P-12          &  $0.73 ^{+0.02 }_{-0.02 }$    &    $0.70 ^{+0.02 }_{-0.01 }$    &    $4650^{+60 }_{-60 }$   &         $-0.29^{+0.05}_{-0.05}$         &    $2.5^{+2.0}_{-2.0}$   \\[0.2cm]
   TrES-4            &  $1.404^{+0.066}_{-0.134}$    &    $1.846^{+0.096}_{-0.087}$    &    $6200^{+75 }_{-75 }$   &         $+0.14^{+0.09}_{-0.09}$         &    $2.9^{+1.5}_{-0.4}$   \\[0.2cm]
   WASP-4            &  $0.925^{+0.040}_{-0.040}$    &    $0.912^{+0.013}_{-0.013}$    &    $5500^{+100}_{-100}$   &         $-0.03^{+0.09}_{-0.09}$         &    $6.5^{+2.3}_{-2.3}$   \\[0.2cm]
   WASP-6            &  $0.880^{+0.050}_{-0.080}$    &    $0.870^{+0.025}_{-0.036}$    &    $5450^{+100}_{-100}$   &         $-0.20^{+0.09}_{-0.09}$         &    $11 ^{+7  }_{-7  }$   \\[0.2cm]
   WASP-12           &  $1.35 ^{+0.14 }_{-0.14 }$    &    $1.57 ^{+0.07 }_{-0.07 }$    &    $6300^{+200}_{-100}$   &         $+0.30^{+0.05}_{-0.15}$         &    $2  ^{+1  }_{-1  }$   \\[0.2cm]
   WASP-15           &  $1.18 ^{+0.12 }_{-0.12 }$    &    $1.477^{+0.072}_{-0.072}$    &    $6300^{+100}_{-100}$   &         $-0.17^{+0.11}_{-0.11}$         &    $3.9^{+2.8}_{-1.3}$   \\[0.2cm]
  \tableline
  \tablenotetext{}{Notes $-$ The references are the same as the ones used for the data in Table {\ref{tab:transit_planets_data}}. The ages are less well constrained
                   and should be taken with caution.}

\end{tabular}
\end{center}
\end{table}

\clearpage
\begin{table}[ht]
\small
\begin{center}
\caption{Derived Power Needed ($M_{\rm core}=0$)} \label{tab:necessary_heating}
\vspace{0.2in}
\begin{tabular}{|l|cc|c|cc|cc|}
\hline
\rule {0pt} {10pt}
\multirow{2}{*}{Planet} &\multicolumn{2} {c|}{$\dot{E}_{\rm heating} (L_{\sun})$} & $\dot{E}_{\rm irradiation} (L_{\sun})$$^a$ & \multicolumn{2} {c|}{$\dot{E}_{\rm heating}/\dot{E}_{\rm irradiation}$}& \multicolumn{2} {c|}{$(e/0.05)^2 / (Q'_p/10^5)$} \\
                        &      solar            &   $10\times$solar               &                                            &        solar           & $10\times$solar                               &     solar            & $10\times$solar         \\
\hline
\rule {0pt} {10pt}
   HAT-P-12b$^b$       &  $       -          $  &   $       -          $          &        $7.4 \times 10^{-6}$                &   $       -          $ & $       -          $                          & $-$                  & $-$                  \\[0.2cm]
   TrES-4              &  $2.1 \times 10^{-6}$  &   $1.4 \times 10^{-7}$          &        $3.2 \times 10^{-4}$                &   $6.6 \times 10^{-3}$ & $4.4 \times 10^{-4}$                          & $1.1 \times 10^{0}$  & $7.5 \times 10^{-2}$ \\[0.2cm]
   WASP-4b             &  $7.8 \times 10^{-8}$  &   $1.1 \times 10^{-8}$          &        $1.3 \times 10^{-4}$                &   $5.8 \times 10^{-4}$ & $8.2 \times 10^{-5}$                          & $1.3 \times 10^{-3}$ & $1.8 \times 10^{-4}$ \\[0.2cm]
   WASP-6b$^c$         &  $4.6 \times 10^{-10}$ &   $ -                $          &        $2.9 \times 10^{-5}$                &   $1.6 \times 10^{-5}$ & $       -          $                          & $1.2 \times 10^{-3}$ & $-$                  \\[0.2cm]
   WASP-12b            &  $2.4 \times 10^{-6}$  &   $1.5 \times 10^{-7}$          &        $1.2 \times 10^{-3}$                &   $2.0 \times 10^{-3}$ & $1.2 \times 10^{-4}$                          & $3.4 \times 10^{-3}$ & $2.1 \times 10^{-4}$ \\[0.2cm]
   WASP-15b$^d$        &  $5.0 \times 10^{-9}$  &   $       0          $          &        $1.5 \times 10^{-4}$                &   $3.4 \times 10^{-5}$ & $       0          $                          & $1.1 \times 10^{-2}$ & $         0        $ \\[0.2cm]
  \tableline
\end{tabular}
\bigskip
\tablenotetext{}{Notes $-$ This is the minimum heating rate to fit the measured radius. A central core would shrink the radius, and require a higher heating rate.}
\tablenotetext{}{$^{a}$ $\dot{E}_{\rm irradiation}$ is the irradiation rate, i.e. the stellar flux $F_{p}$ (see Table~\ref{tab:transit_planets_data}) intercepted by the planet:
                 $\dot{E}_{\rm irradiation}=F_{p}\pi R_p^2$.}
\tablenotetext{}{$^{b}$ HAT-P-12b is the only non-inflated planet in the list; it has a measured radius smaller than the one predicted by the standard theory
                 without a core. It seems to require a heavy-element central core or a heavy-element envelope to shrink its radius.
                (see Section \ref{subsec:compatible_models} and Table~\ref{tab:limits_Mcore_HAT-P-12b_TrES_4_WASP-4b}).}
\tablenotetext{}{$^{c}$ At $10\times$solar, the theoretical radius of WASP-6b is slightly above the best measured value, though within $1\sigma$.}
\tablenotetext{}{$^{d}$ At $10\times$solar, the theoretical radius of WASP-15b can be fit without tidal heating.}
\end{center}
\end{table}

\vspace{2cm}

\begin{table}[ht]
\small
\begin{center}
\caption{Lower limit, upper limit, and best-estimates of the planets' tidal dissipation factors $Q'_p$ ($M_{\rm core}=0$)} \label{tab:limits_Qp}
\vspace{0.2in}
\begin{tabular}{|l|c|c|cc|}
\hline
\rule {0pt} {10pt}
\multirow{2}{*}{Planet} & \multirow{2}{*}{$Q'_p \rm(min)$}  & \multirow{2}{*}{$Q'_p \rm(max)$}  & \multicolumn{2} {c|}{$Q'_p$ (best-estimate)} \\
                        &                                  &                                  &   solar             & $10\times$solar          \\
\hline
\rule {0pt} {10pt}
   TrES-4$^a$           &           $        -        $    &        $6.7 \times 10^{4}$       & $3.6 \times 10^{3}$ & $5.3 \times 10^{4}$  \\[0.2cm]
   WASP-4b$^b$          &           $        -        $    &        $3.1 \times 10^{9}$       & $2.8 \times 10^{8}$ & $2.2 \times 10^{9}$  \\[0.2cm]
   WASP-6b$^c$          &           $2.0 \times 10^{7}$    &        $        -        $       & $9.0 \times 10^{7}$ & $        -        $  \\[0.2cm]
   WASP-12b             &           $1.0 \times 10^{7}$    &        $1.1 \times 10^{9}$       & $2.9 \times 10^{7}$ & $4.6 \times 10^{8}$  \\[0.2cm]
   WASP-15b             &           $3.8 \times 10^{5}$    &        $1.3 \times 10^{9}$       & $9.8 \times 10^{6}$ & $7.7 \times 10^{7}$  \\[0.2cm]
  \tableline
\end{tabular}
\bigskip
\tablenotetext{}{Notes $-$ $Q'_p$(min) and $Q'_p$(max) are the minimum and maximum limits of $Q'_p$, whatever the planet's atmospheric opacity 
                (solar or $10\times$solar), while considering the uncertainties in the measured values of the eccentricity $e$.
                 $Q'_p$ (best-estimate) is defined such that the theoretical planetary radius for this $Q'_p$ fits the best-estimated measured radius
                 at the best-estimated age and eccentricity as measured and compiled in Tables \ref{tab:transit_planets_data} and \ref{tab:host_stars_data}.
                 Most of the values in this table are above the commonly accepted limits. One way to decrease them is to reconsider the measured eccentricities.
                 These are indeed generally poorly constrained. If lower upper limits on $e$ could be demonstrated, they will reduce the required
                 values of the $Q'_ps$. Another possibility is to assume that the planets have central cores. This would result in the shrinking of their radii, that
                 can otherwise be inflated by larger tidal heating resulting from lower $Q'_ps$.}

\tablenotetext{}{$^{a}$ Since there is no lower limit to the eccentricity of TrES-4, we cannot give a lower limit $Q'_p$(min) for this planet. The best-estimate fit
                 is given under the rough assumption that $e\sim 0.01$ for TrES-4 (see Table~\ref{tab:transit_planets_data}).}
\tablenotetext{}{$^{b}$ Same as above, now for WASP-4b. The best fit is given using the upper limit $e=0.096$ for WASP-4b (see Table~\ref{tab:transit_planets_data}).}
\tablenotetext{}{$^{c}$ Within the $\pm 1 \sigma$ limits, the measured radius of WASP-6b can be fitted by the theory without invoking (tidal) heating, at solar and
                at $10\times$solar.  Therefore, there is no need for tidal heating and consequently no upper limit $Q'_p$(max) for this planet. Moreover,
                since the theoretical radius at $10\times$solar is slightly above the best measured value (though within $\pm 1 \sigma$), $Q'_p$
                (best-estimate) is not defined in this case.}
\end{center}
\end{table}

\begin{table}[ht]
\small
\begin{center}
\caption{Derived core mass for various planet heating rates: the Cases
of HAT-P-12{\rm b}, TrES-4, and WASP-4{\rm b}.}
\label{tab:limits_Mcore_HAT-P-12b_TrES_4_WASP-4b}
\vspace{0.2in}
\begin{tabular}{|l|rrrr|c|c|c|cc|}
\hline
\rule {0pt} {10pt}
\multirow{2}{*}{Planet}     & \multicolumn{4} {c|}{$M_{\rm core} (\rm M_{\earth})(\% M_{p})$}              & $\dot{E}_{\rm heating} (L_{\sun})$ & {$\dot{E}_{\rm heating}/\dot{E}_{\rm irradiation}$} & {$(e/0.05)^2 / (Q'_p/10^5)$}    & \multicolumn{2} {c|}{$\rm \log_{10} (Q'_p)$}\\
                            & \multicolumn{2} {c}{solar}       & \multicolumn{2} {c|}{$10\times$solar} &                                        &                                                     &                                   &              &                                \\
\hline
\rule {0pt} {10pt}
\multirow{4}{*}{HAT-P-12b} &     $ 31$ & $(46\%)$              &    $ 36$ & $(54\%)$                       &      $           0      $          &                $0                 $                 &    $       -          $           &     $ -  $   &	   ($e=0.085$)  \\[0.2cm]
                           &     $ 31$ & $(46\%)$              &    $ 37$ & $(55\%)$                       &      $1.0\times 10^{-10}$          &                $1.4 \times 10^{-5}$                 &    $7.2 \times 10^{-4}$           &     $8.60$   &	      $ -$      \\[0.2cm]
                           &     $ 33$ & $(49\%)$              &    $ 40$ & $(60\%)$                       &      $3.0\times 10^{-10}$          &                $4.1 \times 10^{-5}$                 &    $2.2 \times 10^{-3}$           &     $8.12$   &	      $''$      \\[0.2cm]
                           &     $ 36$ & $(54\%)$              &    $ 44$ & $(66\%)$                       &      $6.0\times 10^{-10}$          &                $8.1 \times 10^{-5}$                 &    $4.3 \times 10^{-3}$           &     $7.82$   &	      $''$      \\[0.2cm]
                           &     $ 40$ & $(60\%)$              &    $ 48$ & $(72\%)$                       &      $1.3\times 10^{-9}$	        &                $1.8 \times 10^{-4}$                 &    $9.4 \times 10^{-3}$           &     $7.49$   &	      $''$      \\[0.2cm]
                           &     $ 45$ & $(67\%)$              &    $  -$ & $(  - )$                       &      $3.0\times 10^{-9}$           &                $4.1 \times 10^{-4}$                 &    $2.2 \times 10^{-2}$           &     $7.12$   &	      $''$      \\[0.2cm]
                           &     $ 48$ & $(72\%)$              &    $  -$ & $(  - )$                       &      $5.0\times 10^{-9}$           &                $6.8 \times 10^{-4}$                 &    $3.6 \times 10^{-2}$           &     $6.90$   &	      $''$      \\[0.2cm]
\hline
\rule {0pt} {10pt}
\multirow{4}{*}{TrES-4}    &     $  -$ & $(  - )$              &    $  0$ & $( 0\%)$                       &      $1.4\times 10^{-7}$           &                $4.4 \times 10^{-4}$                 &    $7.5 \times 10^{-2}$           &     $4.72$   &	   ($e=0.01$)   \\[0.2cm]
                           &     $  -$ & $(  - )$              &    $ 15$ & $( 5\%)$                       &      $2.4\times 10^{-7}$           &                $7.6 \times 10^{-4}$                 &    $1.3 \times 10^{-1}$           &     $4.50$   &	      $''$      \\[0.2cm]
                           &     $  -$ & $(  - )$              &    $ 37$ & $(13\%)$                       &      $4.2\times 10^{-7}$           &                $1.3 \times 10^{-3}$                 &    $2.3 \times 10^{-1}$           &     $4.25$   &	      $''$      \\[0.2cm]
                           &     $  -$ & $(  - )$              &    $ 62$ & $(21\%)$                       &      $7.4\times 10^{-7}$           &                $2.3 \times 10^{-3}$                 &    $4.0 \times 10^{-1}$           &     $4.00$   &	      $''$      \\[0.2cm]
                           &     $  -$ & $(  - )$              &    $ 85$ & $(29\%)$                       &      $1.3\times 10^{-6}$           &                $4.1 \times 10^{-3}$                 &    $7.0 \times 10^{-1}$           &     $3.75$   &	      $''$      \\[0.2cm]
                           &     $  0$ & $( 0\%)$              &    $105$ & $(36\%)$                       &      $2.1\times 10^{-6}$           &                $6.6 \times 10^{-3}$                 &    $1.1 \times 10^{0} $           &     $3.55$   &	      $''$      \\[0.2cm]
                           &     $  3$ & $( 1\%)$              &    $111$ & $(38\%)$                       &      $2.4\times 10^{-6}$           &                $7.4 \times 10^{-3}$                 &    $1.3 \times 10^{0} $           &     $3.50$   &	      $''$      \\[0.2cm]
                           &     $ 22$ & $( 7\%)$              &    $142$ & $(48\%)$                       &      $4.2\times 10^{-6}$           &                $1.3 \times 10^{-2}$                 &    $2.2 \times 10^{0} $           &     $3.25$   &	      $''$      \\[0.2cm]
                           &     $ 45$ & $(15\%)$              &    $171$ & $(58\%)$                       &      $7.4\times 10^{-6}$           &                $2.3 \times 10^{-2}$                 &    $4.0 \times 10^{0} $           &     $3.00$   &	      $''$      \\[0.2cm]
\hline
\rule {0pt} {10pt}
\multirow{4}{*}{WASP-4b}   &     $  -$ & $(  - )$              &    $  0$ & $( 0\%)$                       &      $1.1\times 10^{-8}$           &                $8.2 \times 10^{-5}$                 &    $1.8 \times 10^{-4}$           &     $9.32$   &	   ($e=0.096$)  \\[0.2cm]
                           &     $  -$ & $(  - )$              &    $ 26$ & $( 7\%)$                       &      $2.2\times 10^{-8}$           &                $1.6 \times 10^{-4}$                 &    $3.7 \times 10^{-4}$           &     $9.00$   &	      $''$      \\[0.2cm]
                           &     $  -$ & $(  - )$              &    $ 49$ & $(12\%)$                       &      $3.9\times 10^{-8}$           &                $2.9 \times 10^{-4}$                 &    $6.5 \times 10^{-4}$           &     $8.75$   &	      $''$      \\[0.2cm]
                           &     $  0$ & $( 0\%)$              &    $ 78$ & $(20\%)$                       &      $7.8\times 10^{-8}$           &                $5.8 \times 10^{-4}$                 &    $1.3 \times 10^{-3}$           &     $8.45$   &	      $''$      \\[0.2cm]
                           &     $ 26$ & $( 7\%)$              &    $125$ & $(32\%)$                       &      $2.2\times 10^{-7}$           &                $1.6 \times 10^{-3}$                 &    $3.7 \times 10^{-3}$           &     $8.00$   &	      $''$      \\[0.2cm]
                           &     $ 47$ & $(12\%)$              &    $156$ & $(40\%)$                       &      $3.9\times 10^{-7}$           &                $2.9 \times 10^{-3}$                 &    $6.6 \times 10^{-3}$           &     $7.75$   &	      $''$      \\[0.2cm]
                           &     $ 68$ & $(17\%)$              &    $185$ & $(47\%)$                       &      $7.0\times 10^{-7}$           &                $5.2 \times 10^{-3}$                 &    $1.2 \times 10^{-2}$           &     $7.50$   &	      $''$      \\[0.2cm]
                           &     $ 89$ & $(23\%)$              &    $212$ & $(54\%)$                       &      $1.2\times 10^{-6}$           &                $9.3 \times 10^{-3}$                 &    $2.1 \times 10^{-2}$           &     $7.25$   &	      $''$      \\[0.2cm]
                           &     $112$ & $(28\%)$              &    $241$ & $(61\%)$                       &      $2.2\times 10^{-6}$           &                $1.6 \times 10^{-2}$                 &    $3.7 \times 10^{-2}$           &     $7.00$   &	      $''$      \\[0.2cm]
                           &     $142$ & $(36\%)$              &    $272$ & $(69\%)$                       &      $4.0\times 10^{-6}$           &                $2.9 \times 10^{-2}$                 &    $6.6 \times 10^{-2}$           &     $6.75$   &	      $''$      \\[0.2cm]
                           &     $173$ & $(44\%)$              &    $300$ & $(76\%)$                       &      $7.0\times 10^{-6}$           &                $5.2 \times 10^{-2}$                 &    $1.2 \times 10^{-2}$           &     $6.50$   &	      $''$      \\[0.2cm]
                           &     $208$ & $(53\%)$              &    $321$ & $(82\%)$                       &      $1.2\times 10^{-5}$           &                $9.3 \times 10^{-2}$                 &    $2.1 \times 10^{-1}$           &     $6.25$   &	      $''$      \\[0.2cm]
                           &     $247$ & $(63\%)$              &    $  -$ & $(  - )$                       &      $2.2\times 10^{-5}$           &                $1.6 \times 10^{-1}$                 &    $3.7 \times 10^{-1}$           &     $6.00$   &	      $''$      \\[0.2cm]
  \tableline
\end{tabular}
\bigskip
\tablenotetext{}{Notes $-$ The core masses are in $\rm M_{\earth}$ and in percentage of the total mass $\rm M_{p}$ of the planet, at solar and $10\times$solar atmospheric
                 opacities.  The data for the cases with $\rm M_{\rm core}=0$ are the same as in Table~\ref{tab:necessary_heating}. The heating rate,
                 $\dot{E}_{\rm heating} (L_{\sun})$, may have any origin. It is also expressed in terms of the ratio $\dot{E}_{\rm heating}/\dot{E}_{\rm irradiation}$.
		 In case the heating is due to tides, we provide the scaled ratio between the squared orbital eccentricity $e$ and the
                 tidal dissipation factor in the planet $Q'_p$, $(e/0.05)^2 / (Q'_p/10^5)$. Thus, we have a link between a not very well known parameter $e$ and a
                 poorly constrained one $Q'_p$. Still, if in addition we adopt the currently best-estimated values of $e$, as listed in Table~\ref{tab:transit_planets_data}, we
                 have a measure of the values $Q'_p$. Recall that we assume here steady-state core heating.}
\end{center}
\end{table}

\begin{table}[ht]
\small
\begin{center}
\caption{Derived core mass for various planet heating rates: the Cases of WASP-6{\rm b}, WASP-12{\rm b}, and WASP-15{\rm b}.}
\label{tab:limits_Mcore_WASP-6b_WASP-12b_WASP-15b}
\vspace{0.2in}
\begin{tabular}{|l|rrrr|c|c|c|cc|}
\hline
\rule {0pt} {10pt}
\multirow{2}{*}{Planet}   & \multicolumn{4} {c|}{$M_{\rm core} (M_{\earth})(\% M_{p})$}                  & $\dot{E}_{\rm heating} (L_{\sun})$ & {$\dot{E}_{\rm heating}/\dot{E}_{\rm irradiation}$} & {$(e/0.05)^2 / (Q'_p/10^5)$} & \multicolumn{2} {c|}{$\rm \log_{10} (Q'_p)$}\\
                          & \multicolumn{2} {c}{solar}    & \multicolumn{2} {c|}{$10\times$solar}        &                                    &                                                     &                                   &                   &                           \\
\hline
\rule {0pt} {10pt}
\multirow{4}{*}{WASP-6b}  &    $  -$ & $(  - )$           &     $ 6$ & $( 4\%)$                          &      $        0         $          &    $        0         $                             &    $        -         $           &     $ -  $        &        ($e=0.054$)        \\[0.2cm]
                          &    $  -$ & $(  - )$           &     $ 9$ & $( 6\%)$                          &      $1.4\times 10^{-10}$          &    $4.8 \times 10^{-6}$                             &    $3.7 \times 10^{-4}$           &     $8.50$        &           $''$            \\[0.2cm]
                          &    $  0$ & $( 0\%)$           &     $14$ & $( 9\%)$                          &      $4.6\times 10^{-10}$          &    $1.6 \times 10^{-5}$                             &    $1.2 \times 10^{-3}$           &     $8.00$        &           $''$            \\[0.2cm]
                          &    $ 10$ & $( 6\%)$           &     $27$ & $(17\%)$                          &      $1.4\times 10^{-9}$           &    $4.7 \times 10^{-5}$                             &    $3.7 \times 10^{-3}$           &     $7.50$        &           $''$            \\[0.2cm]
                          &    $ 20$ & $(13\%)$           &     $40$ & $(25\%)$                          &      $4.4\times 10^{-9}$           &    $1.5 \times 10^{-4}$                             &    $1.2 \times 10^{-2}$           &     $7.00$        &           $''$            \\[0.2cm]
                          &    $ 34$ & $(21\%)$           &     $58$ & $(36\%)$                          &      $1.4\times 10^{-8}$           &    $4.7 \times 10^{-4}$                             &    $3.7 \times 10^{-2}$           &     $6.50$	&           $''$            \\[0.2cm]
                          &    $ 50$ & $(31\%)$           &     $82$ & $(51\%)$                          &      $4.4\times 10^{-8}$           &    $1.5 \times 10^{-3}$                             &    $1.2 \times 10^{-1}$           &     $6.00$	&           $''$            \\[0.2cm]
                          &    $ 64$ & $(40\%)$           &    $102$ & $(64\%)$                          &      $1.4\times 10^{-7}$           &    $4.7 \times 10^{-3}$                             &    $3.7 \times 10^{-1}$           &     $5.50$	&           $''$            \\[0.2cm]
                          &    $ 74$ & $(46\%)$           &    $116$ & $(73\%)$ 		         &      $2.5\times 10^{-7}$           &    $8.5 \times 10^{-3}$                             &    $6.6 \times 10^{-1}$           &     $5.25$        &           $''$            \\[0.2cm]
                          &    $ 88$ & $(55\%)$           &    $  -$ & $(  - )$                          &      $4.4\times 10^{-7}$           &    $1.5 \times 10^{-2}$                             &    $1.2 \times 10^{0} $           &     $5.00$        &           $''$            \\[0.2cm]
                          &    $102$ & $(64\%)$           &    $  -$ & $(  - )$                          &      $1.4\times 10^{-6}$           &    $4.7 \times 10^{-2}$                             &    $3.7 \times 10^{0} $           &     $4.50$        &           $''$            \\[0.2cm]
\hline
\rule {0pt} {10pt}
\multirow{4}{*}{WASP-12b} &    $  -$ & $(  - )$           &    $  0$ & $( 0\%)$                          &      $1.5\times 10^{-7}$           &    $1.2 \times 10^{-4}$                         &    $2.1 \times 10^{-4}$            &     $8.66$        &        ($e=0.049$)       \\[0.2cm]
                          &    $  -$ & $(  - )$           &    $ 34$ & $( 8\%)$                          &      $2.7\times 10^{-7}$           &    $2.2 \times 10^{-4}$                         &    $3.8 \times 10^{-4}$            &     $8.40$        &           $''$           \\[0.2cm]
                          &    $  -$ & $(  - )$           &    $ 56$ & $(13\%)$                          &      $3.8\times 10^{-7}$           &    $3.1 \times 10^{-4}$                         &    $5.4 \times 10^{-4}$            &     $8.25$        &           $''$           \\[0.2cm]
                          &    $  -$ & $(  - )$           &    $ 98$ & $(22\%)$                          &      $6.7\times 10^{-7}$           &    $5.5 \times 10^{-4}$                         &    $9.5 \times 10^{-4}$            &     $8.00$        &           $''$           \\[0.2cm]
                          &    $  -$ & $(  - )$           &    $145$ & $(32\%)$                          &      $1.2\times 10^{-6}$           &    $9.8 \times 10^{-4}$                         &    $1.7 \times 10^{-3}$            &     $7.75$        &           $''$           \\[0.2cm]
                          &    $  0$ & $( 0\%)$           &    $196$ & $(44\%)$                          &      $2.4\times 10^{-6}$           &    $2.0 \times 10^{-3}$                         &    $3.4 \times 10^{-3}$            &     $7.45$        &           $''$           \\[0.2cm]
                          &    $ 33$ & $( 7\%)$           &    $255$ & $(57\%)$                          &      $4.3\times 10^{-6}$           &    $3.5 \times 10^{-3}$                         &    $6.1 \times 10^{-3}$            &     $7.20$        &           $''$           \\[0.2cm]
                          &    $ 64$ & $(14\%)$           &    $296$ & $(66\%)$                          &      $6.7\times 10^{-6}$           &    $5.5 \times 10^{-3}$                         &    $9.6 \times 10^{-3}$            &     $7.00$        &           $''$           \\[0.2cm]
                          &    $110$ & $(25\%)$           &    $351$ & $(78\%)$                          &      $1.2\times 10^{-5}$           &    $9.8 \times 10^{-3}$                         &    $1.7 \times 10^{-2}$            &     $6.75$        &           $''$           \\[0.2cm]
                          &    $165$ & $(37\%)$           &    $  -$ & $(  - )$                          &      $2.1\times 10^{-5}$           &    $1.7 \times 10^{-2}$                         &    $3.0 \times 10^{-2}$            &     $6.50$        &           $''$           \\[0.2cm]
                          &    $230$ & $(51\%)$           &    $  -$ & $(  - )$                          &      $3.8\times 10^{-5}$           &    $3.1 \times 10^{-2}$                         &    $5.4 \times 10^{-2}$            &     $6.25$        &           $''$           \\[0.2cm]
                          &    $300$ & $(67\%)$           &    $  -$ & $(  - )$                          &      $6.7\times 10^{-5}$           &    $5.5 \times 10^{-2}$                         &    $9.6 \times 10^{-2}$            &     $6.00$        &           $''$           \\[0.2cm]
\hline
\rule {0pt} {10pt}
\multirow{4}{*}{WASP-15b} &    $  -$ & $(  - )$           &     $ 0$ & $( 0\%)$                          &      $       0         $           &    $        0         $                         &    $        -         $            &     $  - $        &        ($e=0.052$)       \\[0.2cm]
                          &    $  -$ & $(  - )$           &     $13$ & $( 8\%)$                          &      $1.6\times 10^{-9}$           &    $1.1 \times 10^{-5}$                         &    $3.4 \times 10^{-3}$            &     $7.50$        &           $''$           \\[0.2cm]
                          &    $  0$ & $( 0\%)$           &     $30$ & $(17\%)$                          &      $5.0\times 10^{-9}$           &    $3.4 \times 10^{-5}$                         &    $1.1 \times 10^{-2}$            &     $7.00$        &           $''$           \\[0.2cm]
                          &    $ 15$ & $( 9\%)$           &     $52$ & $(30\%)$                          &      $1.6\times 10^{-8}$           &    $1.1 \times 10^{-4}$                         &    $3.4 \times 10^{-2}$            &     $6.50$        &           $''$           \\[0.2cm]
                          &    $ 35$ & $(20\%)$           &     $73$ & $(42\%)$                          &      $5.1\times 10^{-8}$           &    $3.5 \times 10^{-4}$                         &    $1.1 \times 10^{-1}$            &     $6.00$        &           $''$           \\[0.2cm]
                          &    $ 47$ & $(27\%)$           &     $93$ & $(54\%)$                          &      $1.6\times 10^{-7}$           &    $1.1 \times 10^{-3}$                         &    $3.4 \times 10^{-1}$            &     $5.50$        &           $''$           \\[0.2cm]
                          &    $ 67$ & $(39\%)$           &    $117$ & $(68\%)$                          &      $5.1\times 10^{-7}$           &    $3.5 \times 10^{-3}$                         &    $1.1 \times 10^{0}$             &     $5.00$        &           $''$           \\[0.2cm]
                          &    $ 76$ & $(44\%)$           &    $125$ & $(73\%)$                          &      $9.1\times 10^{-7}$           &    $6.3 \times 10^{-3}$                         &    $1.9 \times 10^{0}$             &     $4.75$        &           $''$           \\[0.2cm]
                          &    $ 85$ & $(49\%)$		  &    $  -$ & $(  - )$		                 &      $1.6\times 10^{-6}$           &    $1.1 \times 10^{-2}$                         &    $3.4 \times 10^{0}$             &     $4.50$        &           $''$           \\[0.2cm]
                          &    $108$ & $(63\%)$           &    $  -$ & $(  - )$                          &      $5.1\times 10^{-6}$           &    $3.5 \times 10^{-2}$                         &    $1.1 \times 10^{1}$             &     $4.00$        &           $''$           \\[0.2cm]
  \tableline
\end{tabular}
\bigskip
\tablenotetext{}{Notes $-$  Same as Table~\ref{tab:limits_Mcore_HAT-P-12b_TrES_4_WASP-4b}, but for different planets.}
\end{center}
\end{table}

\clearpage
\begin{landscape}
\begin{figure}[ht]
\centerline{
\includegraphics[width=12.0cm,angle=0,clip=true]{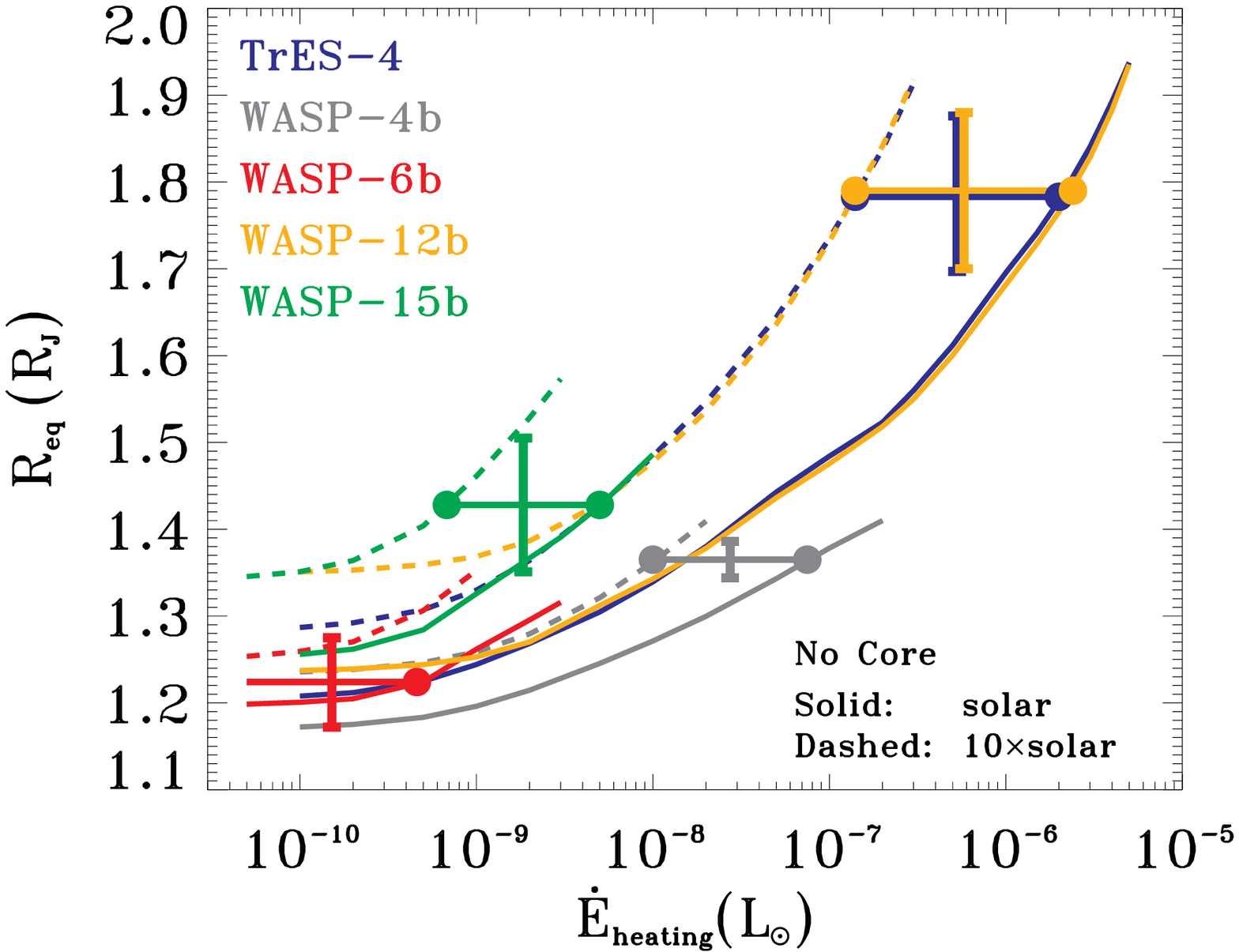}
\includegraphics[width=12.0cm,angle=0,clip=true]{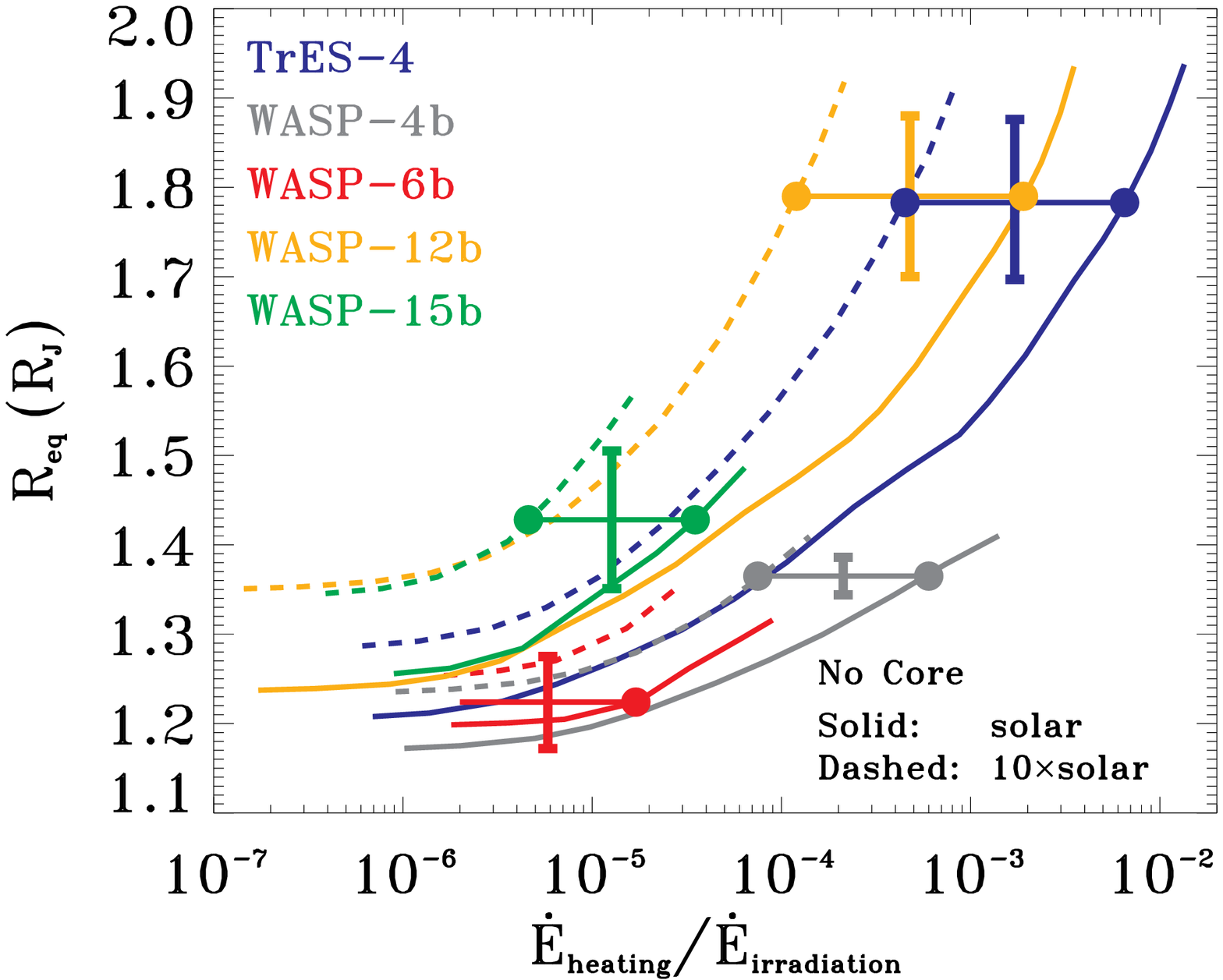}}
\centerline{
\includegraphics[width=12.0cm,angle=0,clip=true]{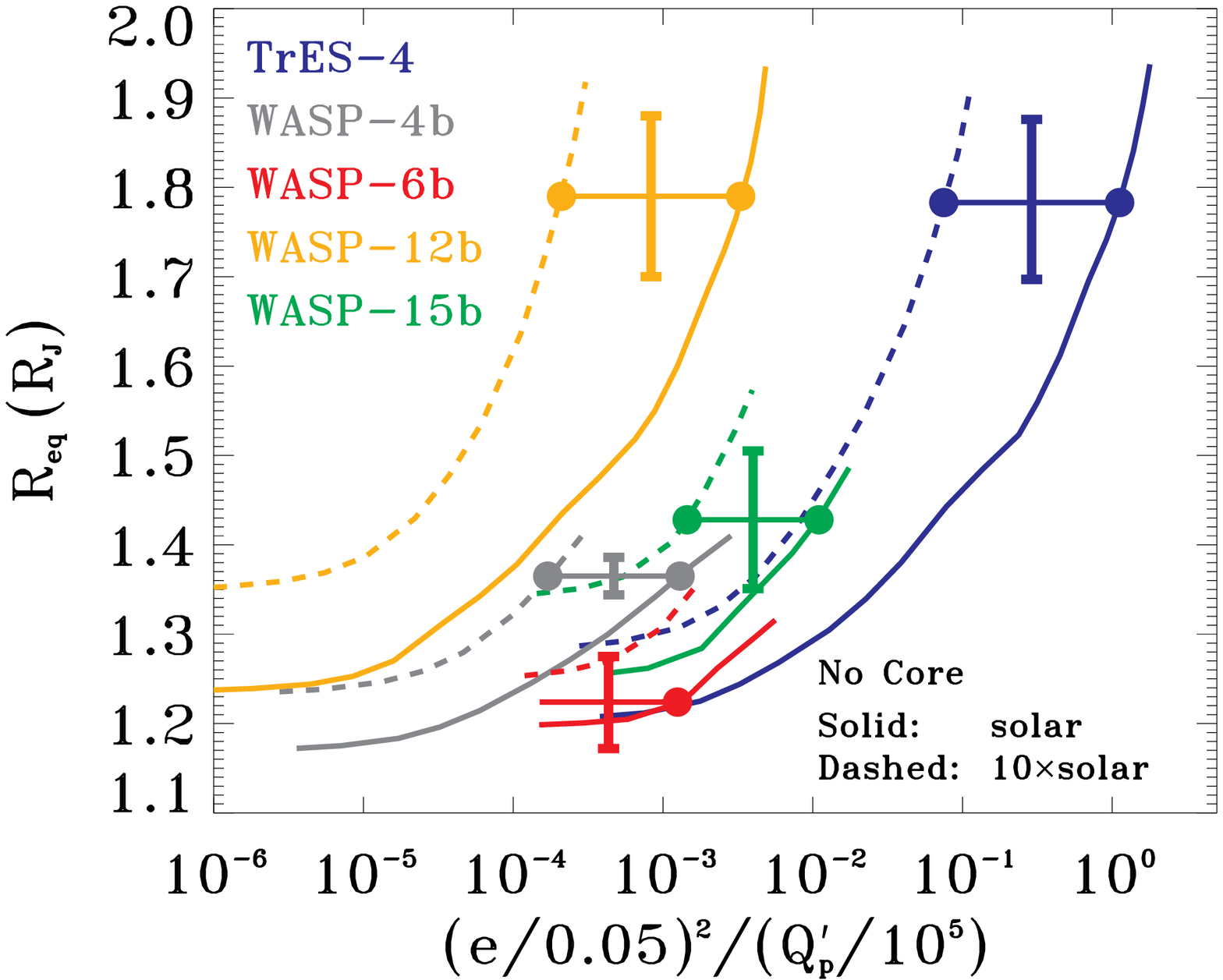}
\includegraphics[width=12.0cm,angle=0,clip=true]{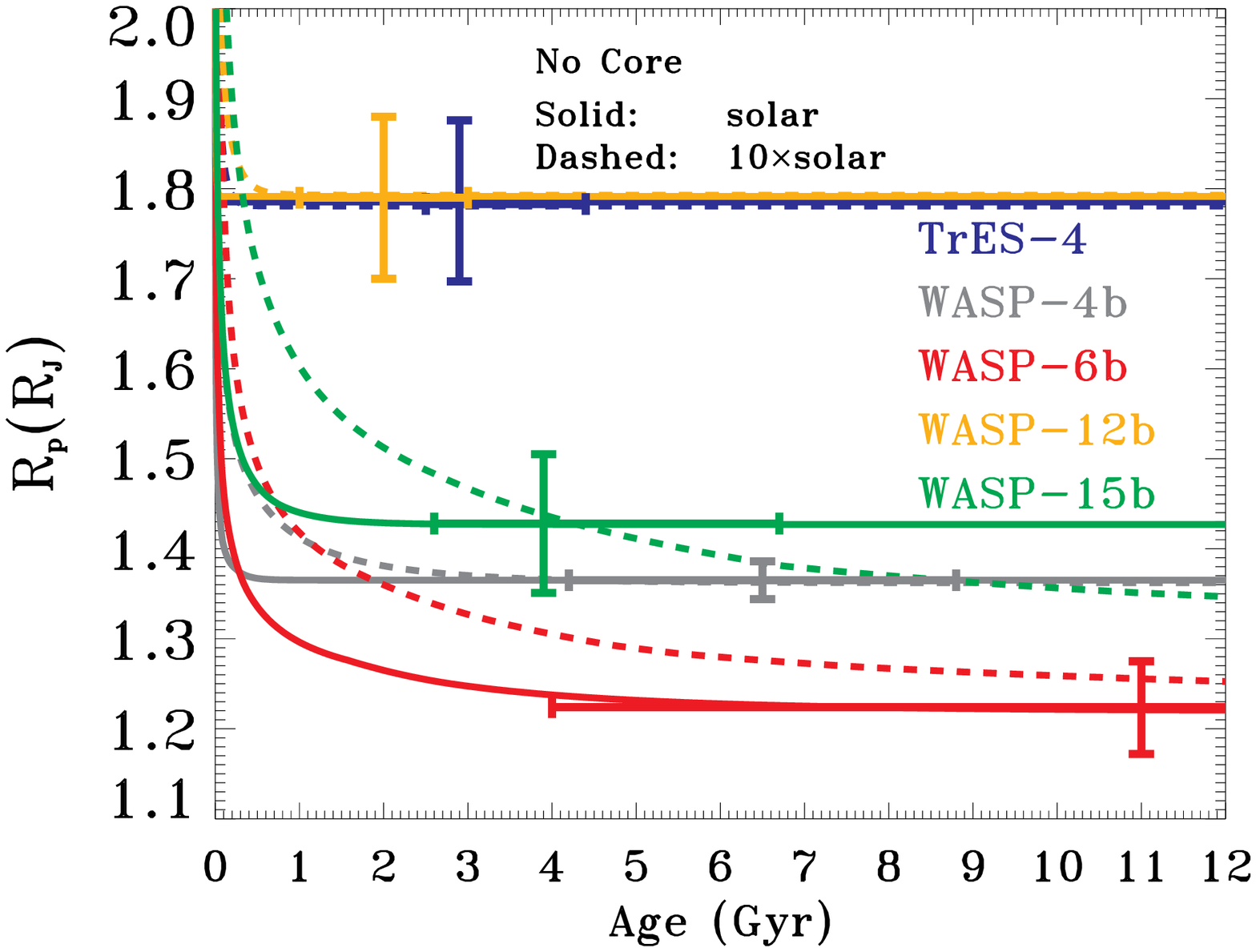}}
\caption{\footnotesize Equilibrium planetary radii $R_{\rm eq}(R_{J})$
          versus the extra-heating rate $\dot{E}_{\rm
          heating}(L_{\sun})$ (top left), the heating rate -
          irradiation rate ratio $\dot{E}_{\rm heating}/\dot{E}_{\rm
          irradiation}$ (top right), and the scaled ratio between the
          squared orbital eccentricity $e$ and the tidal dissipation
          factor in the planet $Q'_p$, $(e/0.05)^2 / (Q'_p/10^5)$
          (bottom left).  The plotted EGPs are TrES-4 (blue), WASP-4b
          (gray), WASP-6b (red), WASP-12b (orange), and WASP-15b
          (green). Here, the planets have no heavy-element central
          core. Two opacities are employed: solar (solid lines), and
          $10\times$solar (dashed lines). The thick horizontal
          segments are the best measured values of the radii and the
          thick vertical segments are the $1\sigma$ tolerances, as
          listed in Table~\ref{tab:transit_planets_data}.  The
          intersections between the best measured values and the
          theoretical $R_{\rm eq}$ are represented by filled circles,
          except for WASP-6b at $10\times$solar opacity, in which case
          this planet is not inflated.  A larger atmospheric opacity
          results in a larger radius, thus in lower necessary internal
          heating (from whatever origin) to fit the measurements.  If
          the heating is due to tides, the bottom left panel provides
          the correlation between the eccentricity $e$ and $Q'_p$.
          Bottom right are depicted the best-fitting theoretical
          radius evolutions versus age (Gyr).  The horizontal segments
          are bounded by the minimum and maximum estimated ages of the
          systems (see Table~\ref{tab:host_stars_data}). At a given
          opacity, the $R_{\rm eq}$ reached by each evolutionary
          curve, and the associated heating, is depicted by one of the
          filled circles in each of the three other panels. The
          example of WASP-15b at $10\times$solar demonstrates that the
          equilibrium radius may not be reached at the age of the
          system, if we assume a extra heating at constant rate. See
          section \ref{subsec:no_central_core} for more discussion.}
\label{fig:fig1}
\end{figure}

\clearpage
\begin{figure}[ht]
\centerline{
\includegraphics[width=12.0cm,angle=0,clip=true]{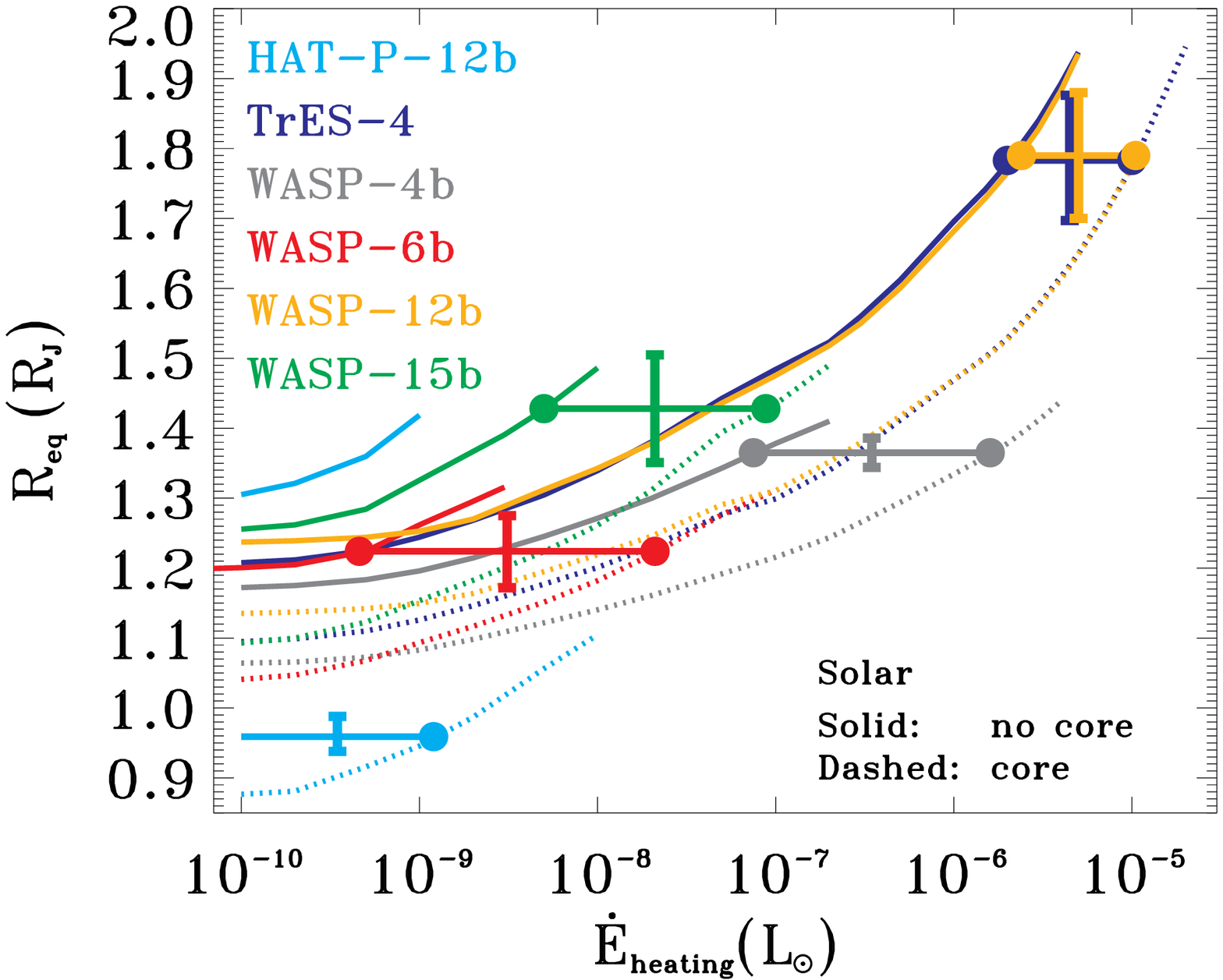}
\includegraphics[width=12.0cm,angle=0,clip=true]{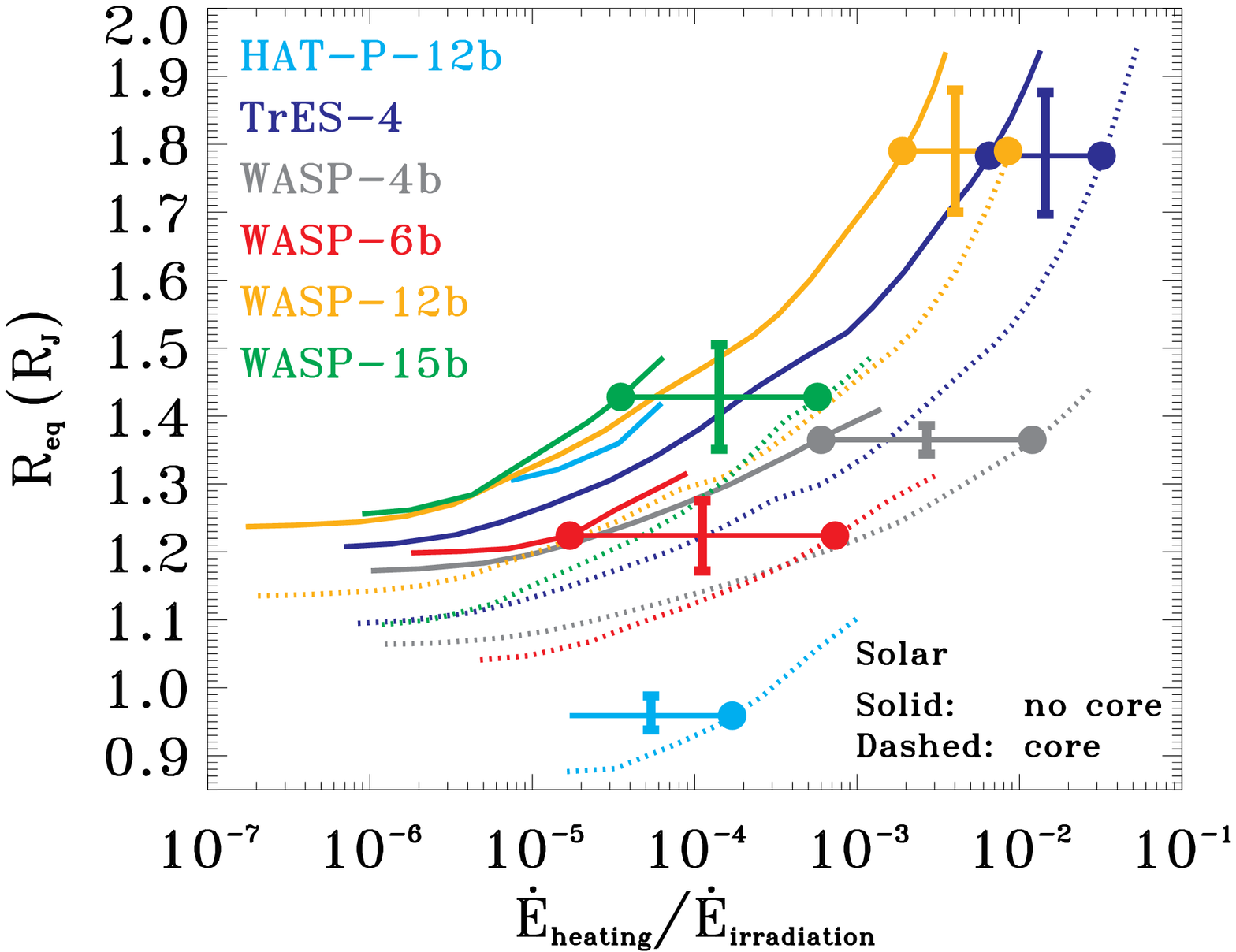}}
\centerline{
\includegraphics[width=12.0cm,angle=0,clip=true]{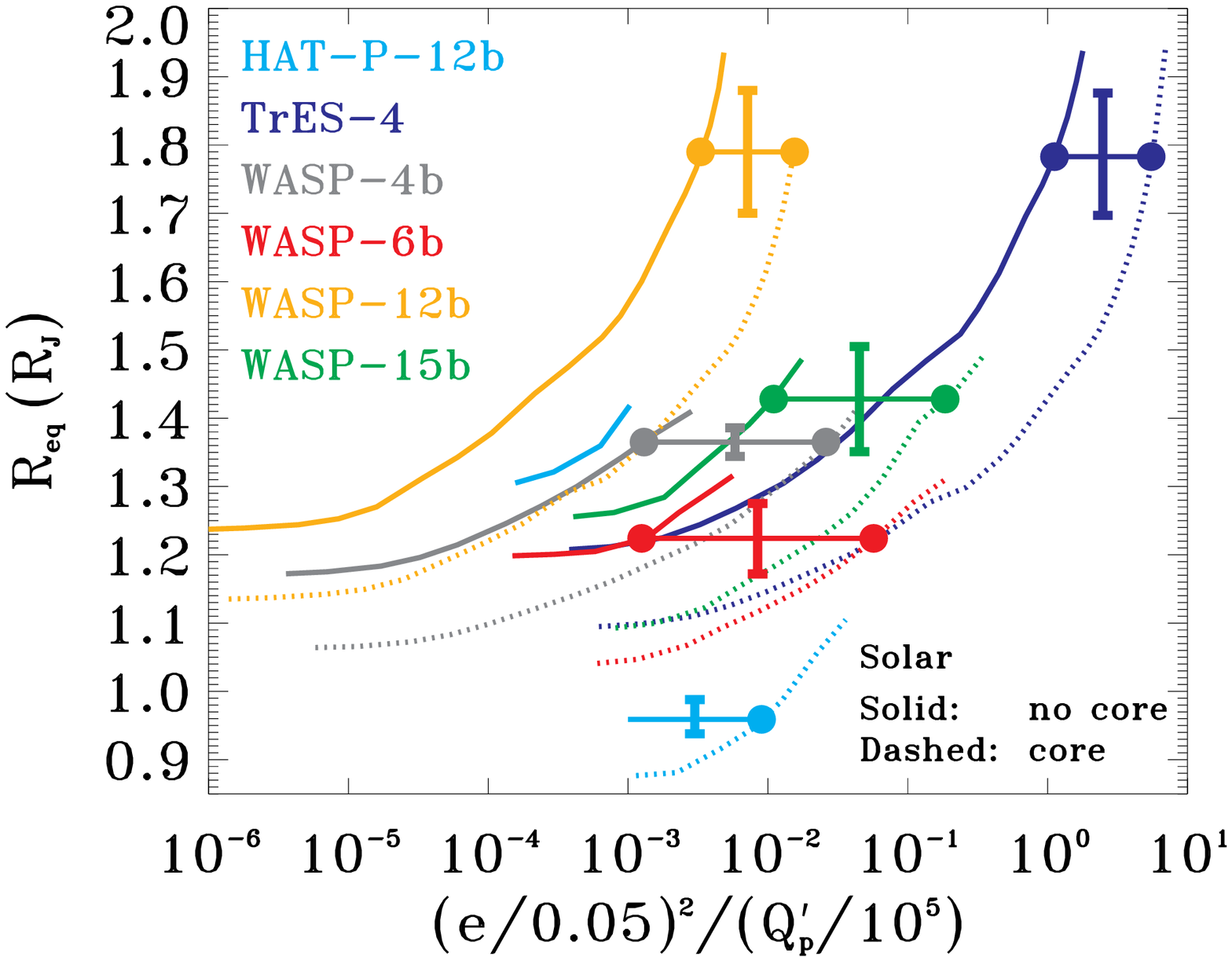}
\includegraphics[width=12.0cm,angle=0,clip=true]{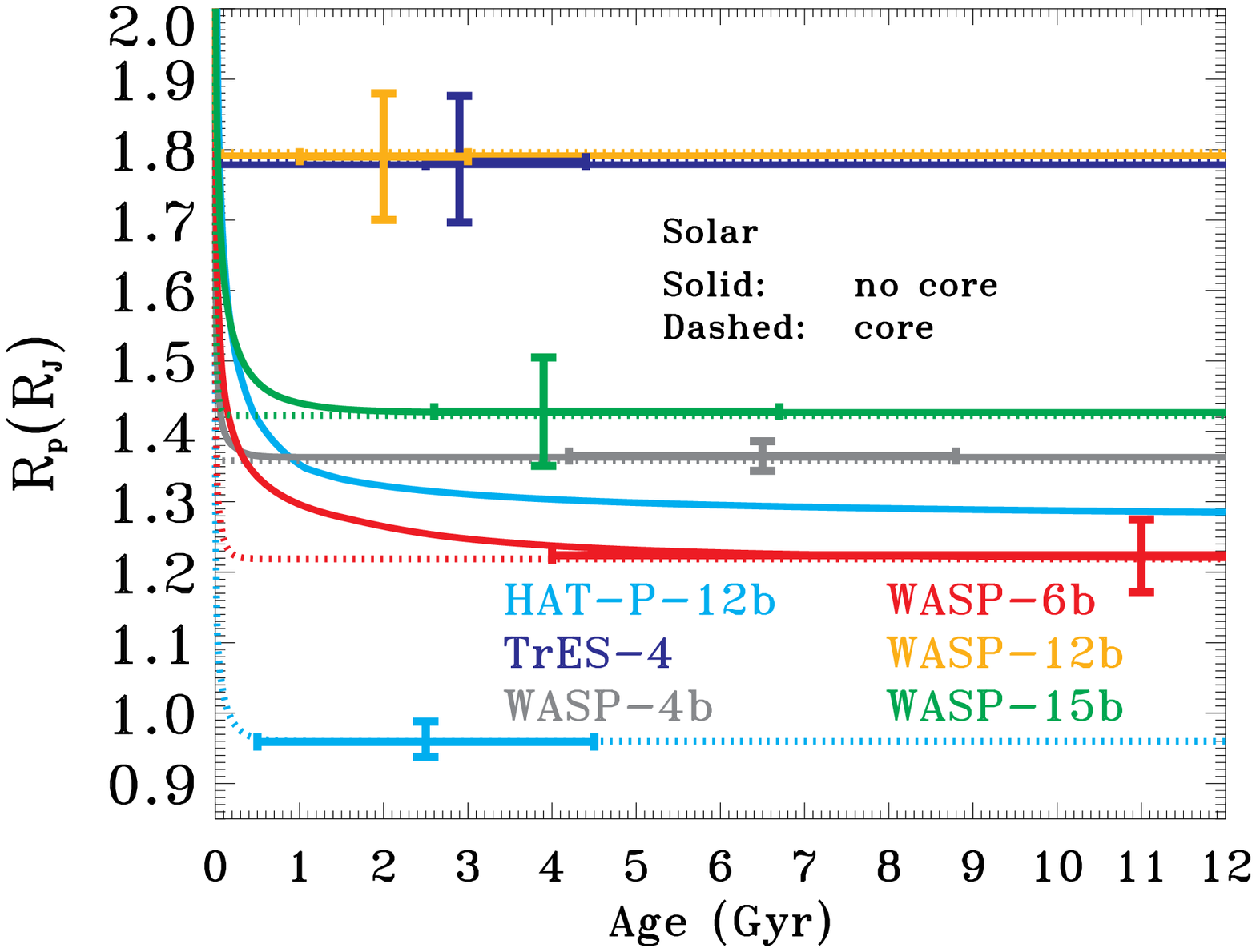}}
\caption{\footnotesize Same as in Figure~\ref{fig:fig1}, but depicting
         the effect of a heavy-element central core at solar opacity.
         HAT-P-12b, not an inflated planet and therefore not
         represented in Figure~\ref{fig:fig1}, is added here because
         it probably has a core. The solid curves represent the cases
         with no central core -- they are identical to the ones in
         Figure~\ref{fig:fig1}.  The dotted curves represent examples
         with heavy-element central cores.  The actual core masses are
         unknown parameters. Ranges of solutions exist, as depicted in
         Fig.~\ref{fig:fig3}, listed in
         Tables~\ref{tab:limits_Mcore_HAT-P-12b_TrES_4_WASP-4b} and
         \ref{tab:limits_Mcore_WASP-6b_WASP-12b_WASP-15b}, and
         explained in Section~\ref{subsec:compatible_models}.
         \textit{For the purpose of illustration only}, we have chosen
         here the following core masses (in $M_{\earth}$): 40 for
         HAT-P-12b, 60 for TrES-4, 100 for WASP-4b, 40 for WASP-6b,
         100 for WASP-12b, and 40 for WASP-15b.  When a heavy-element
         central core is present, the required heating rate is higher
         in order to compensate for the radius shrinking effect of the
         core.  The bottom right panel, depicting the best fitting
         theoretical radius evolutions versus age (Gyr) whose
         equilibrium radii $R_{\rm eq}$ are the thick dots in the
         three other panels, shows that the timescale to reach the
         equilibrium radius is shortened if a core is present. This is
         due to the higher heating rate.  }
\label{fig:fig2}
\end{figure}

\clearpage
\begin{figure}[ht]
\centerline{
\includegraphics[width=12.0cm,angle=0,clip=true]{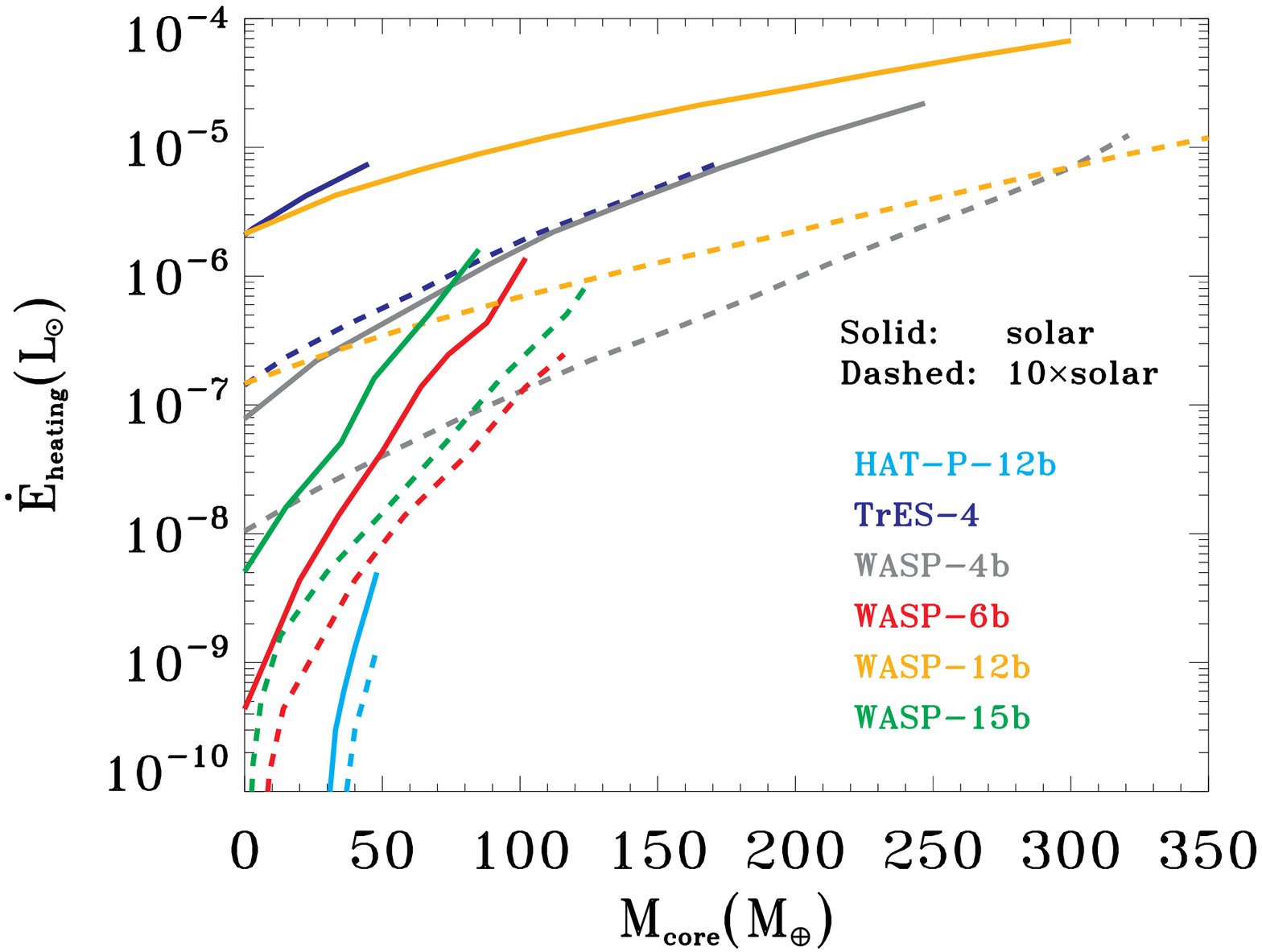}
\includegraphics[width=12.0cm,angle=0,clip=true]{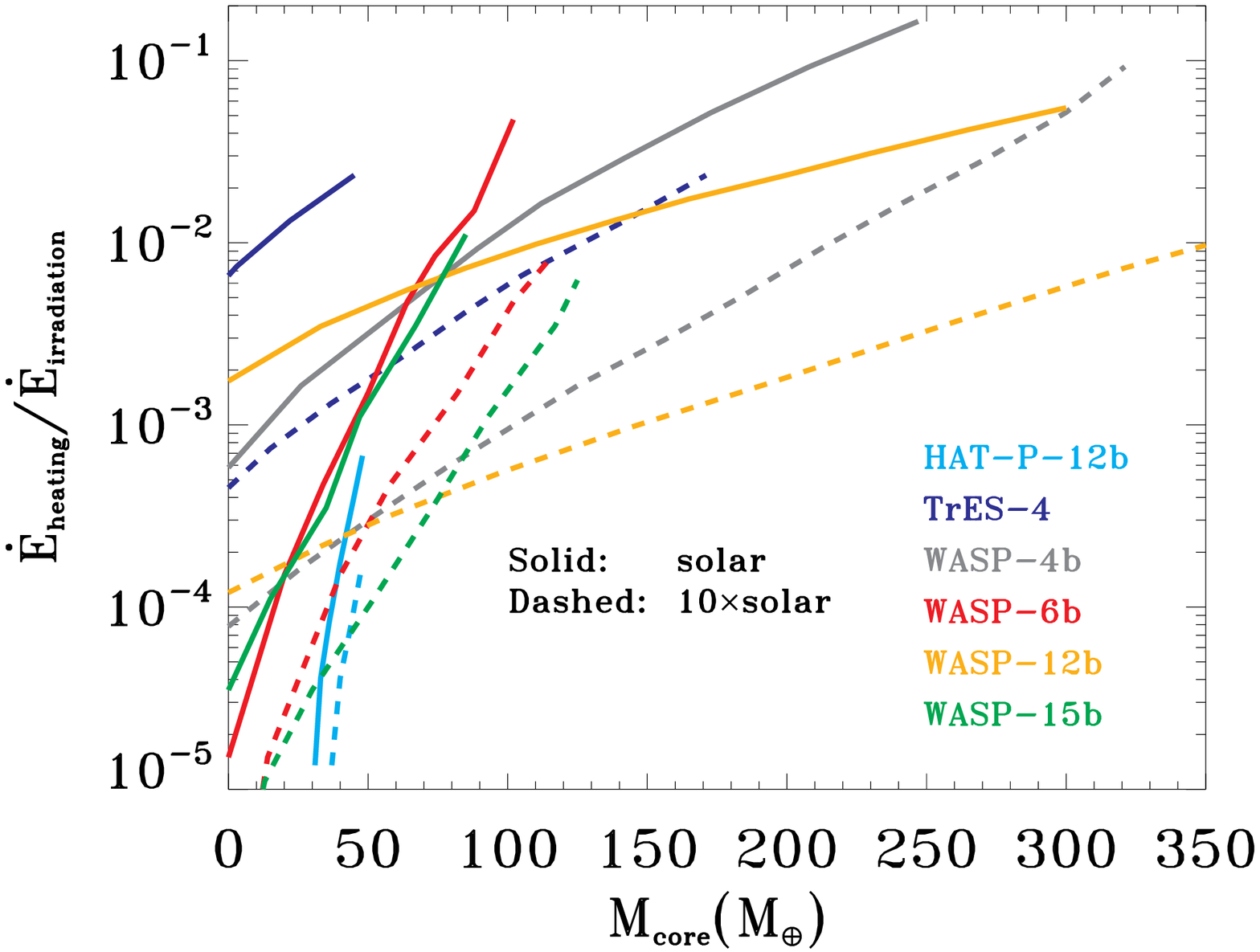}}  
\centerline{
\includegraphics[width=12.0cm,angle=0,clip=true]{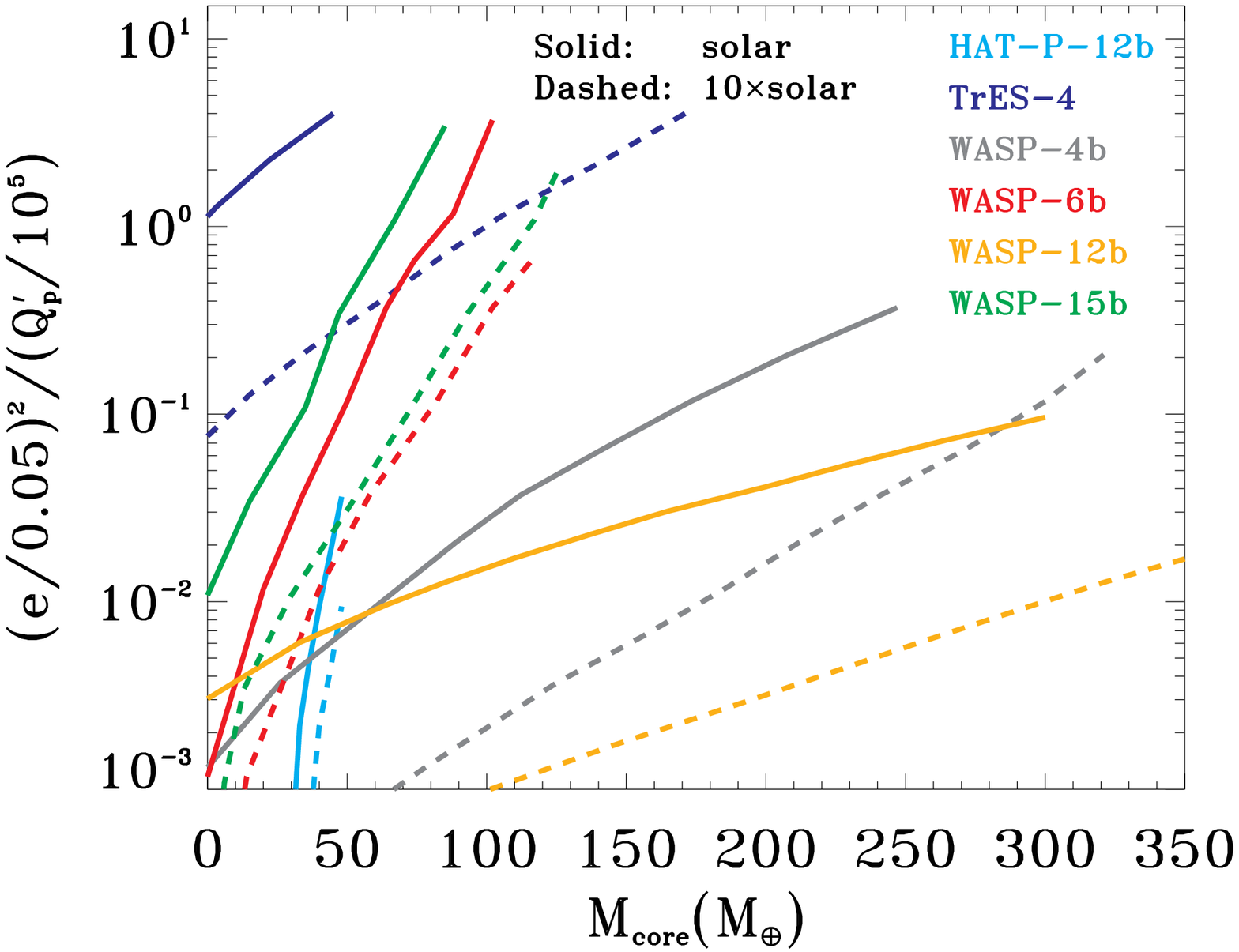}
\includegraphics[width=12.0cm,angle=0,clip=true]{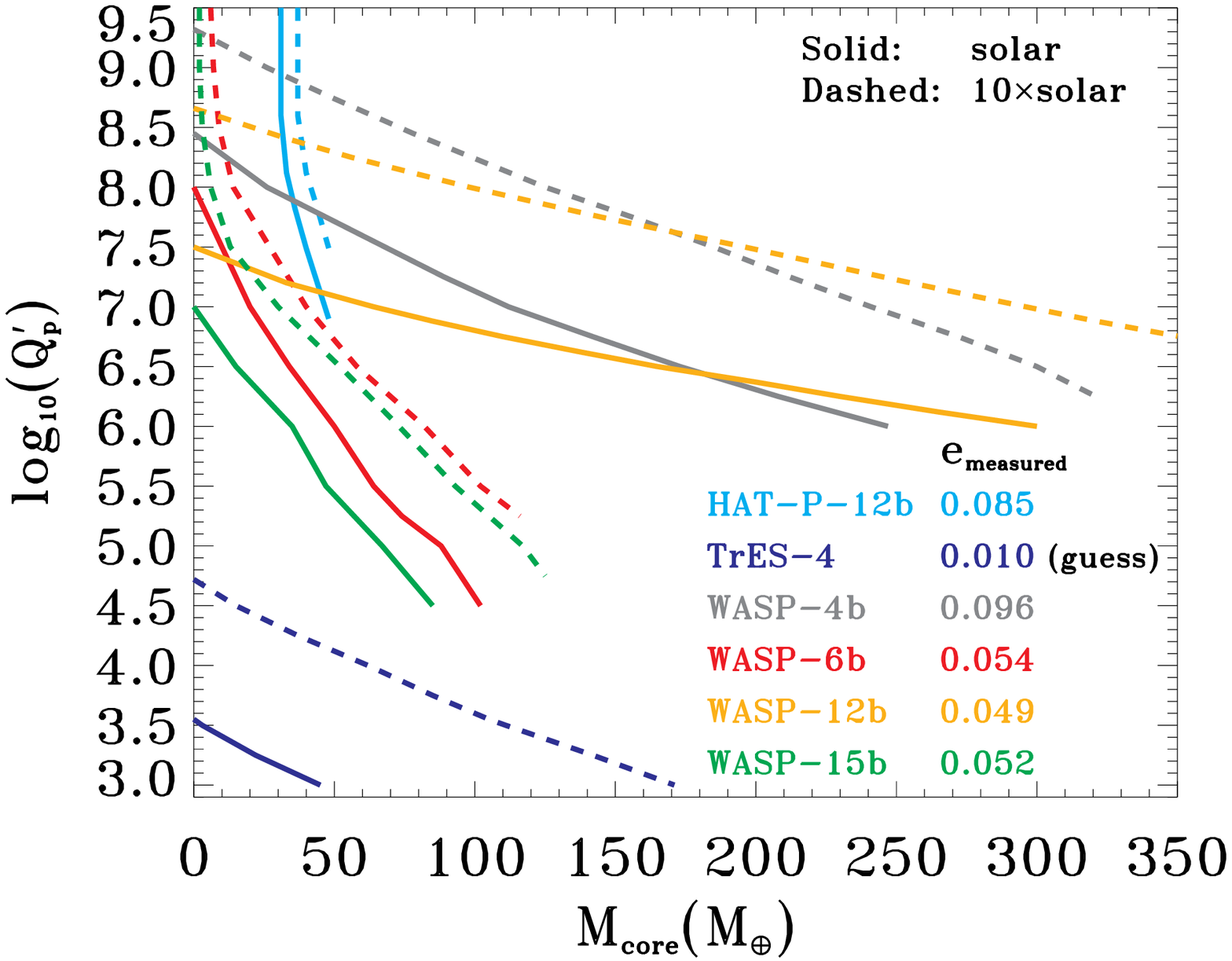}}
\caption{\footnotesize Assuming the steady-state scenario, shown are
         all the possible pairs (heating, $M_{\rm core}
         \left[M_{\earth}\right]$) that enable the theoretical radius
         to fit the measured radius determined within $1\sigma$ for
         each of the following planets: HAT-P-12b (cyan), TrES-4
         (blue), WASP-4b (gray), WASP-6b (red), WASP-12b (orange), and
         WASP-15b (green).  The heating is expressed in terms of
         heating rate $\dot{E}_{\rm heating} (L_{\sun})$ (top left),
         and in terms of the heating rate - irradiation rate ratio,
         $\dot{E}_{\rm heating}/\dot{E}_{\rm irradiation}$ (top
         right). If the heating is due to tides, bottom left 
         panel provides, as a function of $M_{\rm core}$, the scaled ratio
	 between the squared orbital eccentricity $e$ and the tidal 
	 dissipation factor in the planet $Q'_p$, $(e/0.05)^2 / (Q'_p/10^5)$.
	 We provide $\rm log_{10}Q'_p$ if we further make the best-guess 
	 assumption concerning the eccentricities of the planets, listed in the figure and
         in Table~\ref{tab:transit_planets_data}. All these figures
         show that for a given core mass, a larger opacity requires
         less (tidal) heating and, thus, a higher $Q'_p$, everything
         else being equal.}
\label{fig:fig3}
\end{figure}

\clearpage
\end{landscape}

\clearpage


\end{document}